\documentclass[12pt,preprint]{aastex}
\usepackage{epsf}
\usepackage{graphicx}

\newcommand{\etal}{{et al.~}}
\newcommand{\masyr}{ \ {\rm{mas \ yr^{-1}}}\>}
\newcommand{\kms}{ \ {\rm{km \ s^{-1}}}\>}

\begin{document}

\pagenumbering{arabic}
\title{Is the SMC Bound to the LMC? The $HST$ Proper Motion of the SMC}
\author{Nitya Kallivayalil\altaffilmark{1}}
\affil{Harvard-Smithsonian Center for Astrophysics, 
60 Garden Street, Cambridge, MA 02138}
\altaffiltext{1}{nkalliva@cfa.harvard.edu}
\author{Roeland P. van der Marel\altaffilmark{2}}
\affil{Space Telescope Science Institute, 3700 San Martin Drive, Baltimore, 
MD 21218}
\altaffiltext{2}{marel@stsci.edu}
\author{Charles Alcock\altaffilmark{3}}
\affil{Harvard-Smithsonian Center for Astrophysics, 60 Garden Street, 
Cambridge, MA 02138}
\altaffiltext{3}{calcock@cfa.harvard.edu}

\begin{abstract}
We present a measurement of the systemic proper motion of the Small
Magellanic Cloud (SMC) made using the Advanced Camera for Surveys
(ACS) on the \textit{Hubble Space Telescope} (\textit{HST}). We
tracked the SMC's motion relative to 4 background QSOs over a baseline
of approximately 2 years. The measured proper motion is : $\mu_W =
-1.16 \pm 0.18 \masyr, \ \mu_N = -1.17 \pm 0.18 \masyr$. This is the
best measurement yet of the SMC's proper motion. We combine this new
result with our prior estimate of the proper motion of the Large
Magellanic Cloud (LMC) from the same observing program to investigate
the orbital evolution of both Clouds over the past 9 Gyr.  The current
relative velocity between the Clouds is $105 \pm 42 \kms$. Our
investigations of the past orbital motions of the Clouds in a simple
model for the dark halo of the Milky Way imply that the Clouds could
be unbound from each other. However, our data are also consistent with
orbits in which the Clouds have been bound to each other for
approximately a Hubble time. Smaller proper motion errors and better
understanding of the LMC and SMC masses would be required to constrain
their past orbital history and their bound vs.~unbound nature
unambiguously. The new proper motion measurements should be sufficient
to allow the construction of improved models for the origin and
properties of the Magellanic Stream. In turn, this will provide new
constraints on the properties of the Milky Way dark halo.

\end{abstract}

\keywords{galaxies: kinematics and dynamics --  galaxies: interactions
-- Magellanic Clouds}

\section{Introduction}

The Large and Small Magellanic Clouds (LMC \& SMC), both satellites of
the Milky Way (MW), provide a unique opportunity to study 
interacting galaxy pairs and three-body systems. It is expected from
current models of hierarchical structure formation (e.g. Zentner \&
Bullock 2003) that the interaction between the Clouds and the MW will
have played an important role in the dynamical evolution of the MW's
outer parts (e.g. Font \etal 2006). A resonant interaction between the
LMC \& the MW is thought to be responsible for the MW warp (Weinberg
\& Blitz 2006). In return, the MW has had a major influence on the
Clouds' development, including their star formation history (Holtzman
\etal 1997; Harris \& Zaritsky 2001; Smecker-Hane \etal 2002),
structural and chemical evolution (Mathewson \etal 1986; Bekki \&
Chiba 2005), and kinematics (Hatzidimitriou \etal 1993; Cole \etal
2005).  This three-body interaction is also linked to the origin of
the Magellanic Stream, an approximately circum-polar HI feature
that trails the Clouds in their orbit around the MW (Wannier \&
Wrixon, 1972; Mathewson \etal 1974; Putman \etal 1998), the
inter-Cloud bridge (Putman \etal 1998), and the complicated geometry
(Caldwell \& Coulson 1986; Crowl \etal 2001; but see also Welch \etal
1987) and gas distribution of both Clouds (Stanimirovi{\'c} \etal 2004;
Gardiner \etal 1994).  Some current dwarf galaxies and globular
clusters may have originated from tidal stripping as the Clouds
orbited the Galaxy (Lin \etal 1995).

The Magellanic Stream and any stars that have been tidally stripped
from the Clouds as they orbit the MW provide a ``fossil record'' of
the history of the build-up of MW mass (e.g. Belokurov \etal 2006;
Pe{\~n}arrubia \etal 2005; Johnston \etal 1999). Decoding this record
requires detailed modeling, describing how the internal evolution of
the satellites is affected by tides, and sensitive observations that
make it possible to falsify theoretical predictions. A major
uncertainty in this effort is the orbital motion of the Clouds. While
the radial velocities of the Clouds have been measured to high
precision (van der Marel \etal 2002; Harris \& Zaritsky 2006), the
velocity transverse to the line of sight (the proper motion) has been
harder to constrain. This uncertainty complicates the efforts to infer
the fossil record because the past history of the Clouds is ambiguous.

There have been several groups involved in modeling the orbits of the
Clouds around the Galaxy with the intent of reproducing the Magellanic
Stream (Murai \& Fujimoto 1980; Lin \& Lynden-Bell 1982; Heller \&
Rohlfs 1994; Moore \& Davis 1994; Lin \etal 1995; Gardiner \& Noguchi
1996; Yoshizawa \& Noguchi 2003; Mastropietro \etal 2005; Connors
\etal 2005).  The models fall into two main categories: tidal models
and ram-pressure stripping models. The former deal with the tidal
force exerted by the Galaxy on the MCs.  The Stream is modeled as a
product of the tidal disruption of the SMC, and the inter-Cloud region
is the result of a close encounter between the Clouds, which are
assumed to have been a bound system for the past Hubble time. In
general, the conclusions of these studies are that the MCs are near
perigalacticon, they are gravitationally bound to the Galaxy with
apogalacticon beyond 100 kpc, and it is expected that they will become
separated in the next $1$ - $2$ Gyr. The gas in the Stream is thought
to have originated from the SMC between $1$ - $2$ Gyr ago, and the
mass in the Stream is comparable to the gas content in the SMC itself
(Lin \etal 1995, Putman \etal 2003).

The ram-pressure models also invoke an encounter between the Clouds to
produce the inter-Cloud region, and subsequently the Stream is
produced from collisions between the inter-Cloud region gas and either
high velocity clouds in the Galactic halo (`discrete ram-pressure
stripping'; Wayte 1991) or an extended halo of diffuse ionized gas
(`diffuse ram-pressure stripping'; Moore \& Davis 1994).  The
ram-pressure stripping models have some difficulty in producing the
leading arm of the Stream, while the tidal models do so naturally
(Connors \etal 2005). Also, the number density of high velocity clouds
in the outer halo is uncertain, as is the existence of a sufficiently
dense extended gaseous halo. The ram-pressure stripping models have
become quite sophisticated (Mastropietro \etal 2005), but so far only
include the LMC in the analysis and not the SMC. The tidal models too,
while increasingly detailed, have a few open questions which might
only be adequately addressed with the inclusion of more detailed
gas-dynamical properties. For instance, the lack of symmetry between
the leading and trailing arms of the Stream are indicative of drag on
the HI gas, and the apparent lack of stars in the Stream is still
poorly understood. While there is some evidence for tidally stripped
Magellanic stars far ($22\degr$) from the LMC center and in the
direction of the Carina dwarf (Mu{\~n}oz \etal 2006), no stars have
yet been associated with the Magellanic Stream itself.


If the space velocities of both Clouds were known accurately enough
they would be a valuable tool in understanding evolutionary features
of both the Clouds and the MW.  As stated above, the radial velocities
of the Clouds have been more readily determined than the transverse
velocities, which can only be estimated via proper motions. Since the
LMC is closer and larger, there has been much more detailed work on
its proper motion. There is now good general agreement between the
results from various teams (see Kallivayalil \etal 2006; hereafter
Paper~I). But it is worth noting that while the implied LMC space
velocities agree with some of the models of the Stream (Heller \&
Rohlfs 1994), they differ by $\sim100\kms$ from others (Gardiner \&
Noguchi 1996). The proper motion of the SMC, has been much harder to
constrain. A good inertial reference frame for such a measurement has
been hard to come by.  CCD astrometry with a good sample of background
QSOs is necessary to get a handle on its proper motion.


Previous work on the SMC proper motion includes: Kroupa \& Bastian
(1997) who used Hipparcos measurements of 11 stars to get $\mu_W=
-1.23 \pm 0.84, \mu_N=-1.21 \pm 0.75 \masyr$ (here we define the
proper motions $\mu_W, \mu_N$ in the directions west and north as
$\mu_W \equiv -\mu_\alpha \cos(\delta)$ and $\mu_N \equiv
\mu_\delta$). There is also a study by Irwin \etal (1996) using AAT
(Anglo-Australian Telescope) and CTIO (Cerro Tololo Inter-American
Observatory) 4m photographic plates covering a baseline of 15 - 20
years. A measurement of $\mu_W =- 0.92 \pm 0.2, \ \mu_N=-0.69 \pm
0.2\masyr$ is quoted in Irwin (1999), but the analysis of these data
is unpublished. Anderson \& King (2004b) measured a very accurate
relative proper motion between the SMC and the Globular Cluster 47
\textit{Tucanae} of $\Delta \mu_W = -4.716 \pm 0.035 \masyr$ and
$\Delta \mu_N = -1.357 \pm 0.021 \masyr$. When combined with an
estimate of the absolute proper motion of 47 \textit{Tucanae} by
Freire \etal (2003), who report $\mu_W = -5.3 \pm 0.6 \masyr$ and
$\mu_N = -3.3 \pm 0.6 \masyr$, this implies that the SMC's proper
motion is $\mu_W = -0.6 \pm 0.6\masyr$, $\mu_N = -1.9 \pm 0.6
\masyr$. The unweighted average of these three independent SMC proper
motion measurements\footnote{Momany \& Zaggia (2005) recently obtained
  a very different proper motion for the SMC using the USNO CCD
  Astrograph all-sky Catalog (UCAC2) : $\mu_W =-4.44 \masyr; \ \mu_N =
  -2.94 \masyr$. This is inconsistent with our current knowledge of
  the MC-MW system and probably indicates the presence of systematic
  errors in the catalog, as the authors themselves point out. We
  therefore ignore this measurement in the following.} is $\langle
\mu_W \rangle =-0.91 \pm 0.19 \masyr, \langle \mu_N \rangle = -1.28 \pm
0.36\masyr$.  This is broadly consistent with the current
understanding of the Magellanic Stream and the MC-MW system, but the
errors are are not accurate enough to significantly constrain its
dynamics.

In this paper, we present the results of a project that has allowed us
to measure the systemic proper motion of the SMC to 15\% accuracy
using $HST$ observations of a sample of background QSOs, with just a
two year baseline. Our results for the LMC from the same $HST$ program
were presented in Paper~I. \S~2 describes the QSO sample; \S~3
summarizes the analysis of the data; \S~4 presents the SMC proper
motion results. We then move on to an investigation of the global
dynamics of both the LMC and the SMC given our new velocity
measurements. \S~5 describes the Clouds' movements within a fiducial
model of the Galactic halo and \S~6 gives a discussion and summary of
the results.

\section{Sample}

\begin{figure}
\plotone{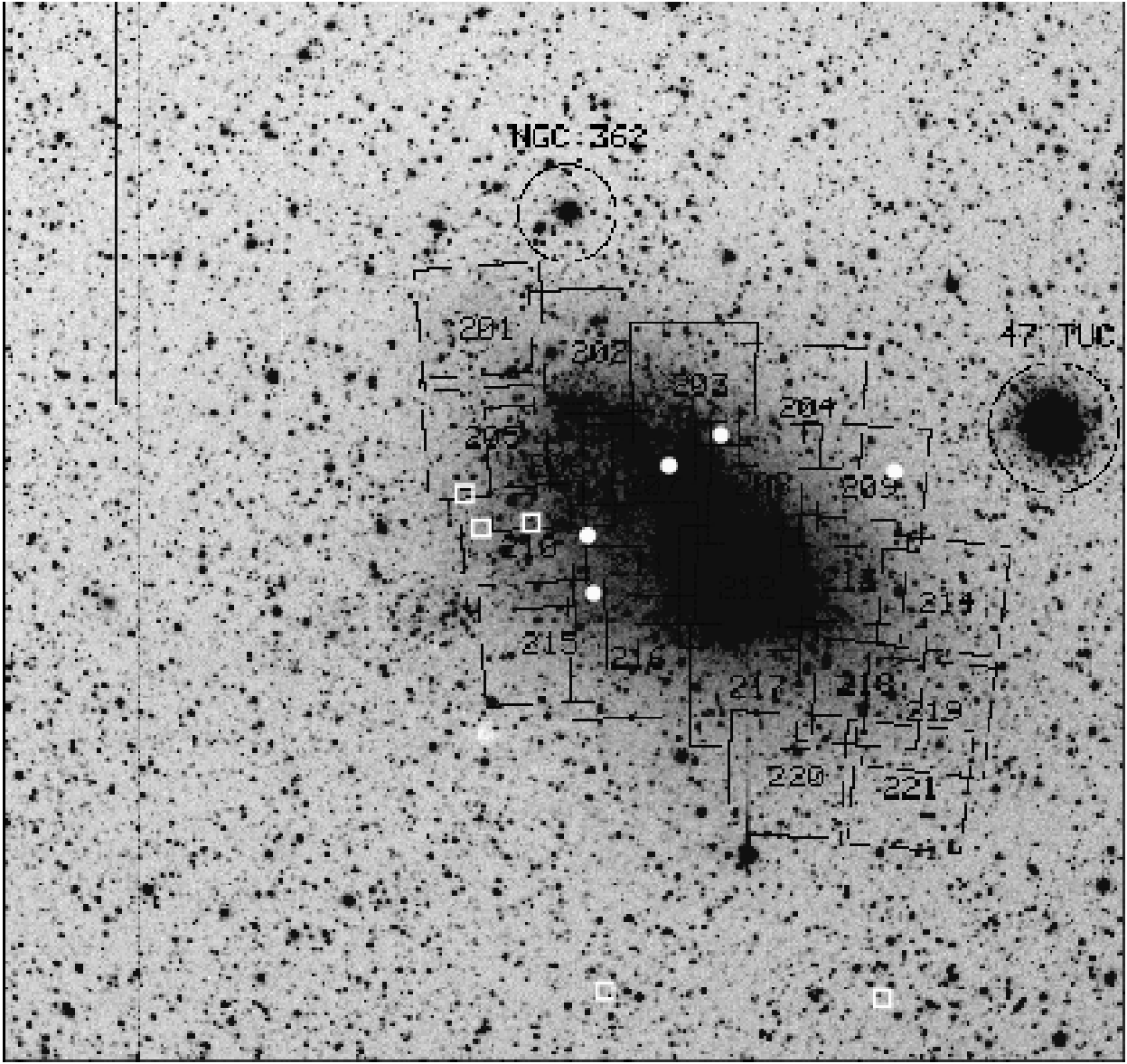}
\caption{R-band image of the SMC ($4^{\circ} \times 4^{\circ}$). 
The MACHO photometric coverage
is indicated (black boxes). White circles indicate reference QSOs for which we
obtained two epochs of ACS/HRC imaging, squares indicate QSOs 
which we did propose for 
but for which we did not get two epochs of imaging in our snapshot program.}
\label{figure1}
\end{figure}

Geha \etal (2003) identified 10 QSOs behind the SMC from their optical
variability in the MACHO database. We proposed to image all 10 fields
using snapshot mode with the High Resolution Camera (HRC) on the
Advanced Camera for Surveys (ACS) on $HST$.  In snapshot mode every
target is not guaranteed, but rather, observations are taken
throughout the cycle, whenever they can be fit in, according to
manually assigned priorities. In the first epoch snapshot program
(Cycle 11; GO 9462) we successfully imaged the fields around 6 of the
10 SMC QSOs.  This is a typical completion rate for snapshot programs.
In the second epoch (Cycle 13; GO 10130) we successfully imaged 5 out
of the 6 proposed SMC QSOs. The 5 QSOs are favorably placed behind the
central few degrees of the SMC where we do not expect to be unduly
influenced by any tidal features (see Kroupa \& Bastian
1997). Figure~\ref{figure1} shows the sample of QSOs behind the
SMC. White circles represent the 5 QSOs for which we did get 2 epochs
of HRC data and white squares mark the QSOs for which we did
not. Table~1 lists the QSO ID, the MACHO ID (for reference with Geha
\etal 2003), the RA, DEC (J2000), $V$-magnitude and redshift for each
of the 5 QSOs.

Our observational strategy is described in detail in Paper~I and the
reader is referred to it for the specifics. The SMC QSOs were part of
the same dataset as the LMC QSOs and the data were taken in an
identical fashion. The HRC was chosen for its high resolution, the
fact that it is well sampled, well calibrated, and that its higher
order distortion has been understood and characterized (Anderson \&
King 2004a; Krist 2003). Our observing strategy was to use many
short-exposure, dithered $V$-band images for the astrometry. In the
first epoch, we imaged each QSO field with the F606W filter (broad
$V$) using two four-point dither patterns that were shifted relatively
by 8 integer pixels (8 frames in total; see Figure~3 in Paper~I for a
schematic of the dither pattern.).  Each field was also imaged with
the F814W filter (broad $I$) using a simple CR-SPLIT. In epoch 2 we
implemented the same F606W sequence, but did not re-do the F814W
observations, since we could use the existing first epoch $I$-band
images for any non-astrometric purposes.  Table~1 lists the date of
observation and exposure times for each filter for each epoch. The
average baseline achieved is $\sim2$ years.

One important potential source of systematic error in these
measurements arises from the relative orientation of the focal plane
at the two epochs. For observations with $HST$, the date of
observation determines the orientation of the detector axes with
respect to the sky. In the case of Paper~I we were in the happy
situation that we had many QSO fields all of which were observed at
random times of the year with random roll angles of the telescope. The
fact that our LMC observations were all carried out at different times
and with different orientations allowed any systematic errors that
were tied to the CCD frame to average out as $\sim1/\sqrt{N}$. We are
not so fortunate with the SMC fields however, because there are far
fewer of them, and because three of them (S1, S2 \& S3) were observed
at about the same time and hence at very similar roll angles.  Table~1
lists the dates of observation in each epoch and the value of
ORIENTAT, which is the position angle on the sky of the detector
$y$-axis, for each of the 5 SMC fields. So if there are systematic
errors that are tied to the CCD frame then they would not have
averaged out to the same extent as for the LMC data. \S~3.2 will
discuss the possibility of systematic errors in detail.

\section{Analysis}
\subsection{Procedure to Obtain a Proper Motion from Each QSO field}
We followed the same procedure to analyze the SMC QSO fields as in
Paper~I (see Figure~4 of Paper~I). We will only summarize our method
here and the interested reader should look to Paper~I for a detailed
explanation. We used Anderson \& King (2004a; hereafter AK04) software
to fit a filter-dependent point-spread function (PSF) to the
flat-fielded images (\_flt.fits) from the STScI pipeline, and to
geometrically correct the raw positions and fluxes of point sources on
the detector chip to account for the higher-order distortion on the
ACS. The geometric distortion is very well understood and has been
characterized by Anderson \& King (AK04) to $\sim 0.005$ pixels. We
then created a master-list of sources for each QSO field by
cross-referencing all 18 frames (16 $V$-band and 2 $I$-band) with the
first $V$-band frame from the first epoch (our `reference' image). In
this way, the master-list only contained sources that were present in
all 18 frames.  We then used a six-parameter linear fit to bring the
star-field of each of the 16 $V$-band frames into alignment with the
reference frame.  This fit accounts for a rotation, a translation, for
scale changes and for skew terms (which account for non-orthogonality
between the two detector axes). These linear transformations are
needed to account for the translational dithering of the observations,
the difference in telescope orientation between epochs, and the
effects of ``breathing'' and differential velocity aberration
(AK04). The linear fit also accounts to lowest order for the
charge-transfer efficiency (CTE) degradation of the telescope (this is
a linear effect with position on the $y$-axis, although in a strict
sense it also depends on stellar brightness). The QSO was removed from
the master-list before this fit was implemented.

Once we had a good working guess at the transformation, cuts were made
in the proper motion (PM)- and proper motion error ($\delta$PM)-space
of the stars in the field to ensure that foreground (Galactic) stars
with large proper motions or stars with large centroiding errors were
not used in the fit. A new six-parameter linear transformation was
found for this star-list, and the process was iterated until the
number of stars used in the fit was unchanging. Once the terms of the
transformations had converged, the reflex motion of the QSO, i.e. the
difference between its average position in the two epochs, was
measured with respect to the star-field using the final
transformation.  This is then $-1\times$(the PM of the SMC). The
error in the SMC PM of each QSO field is estimated as the quadrature
sum of the following two quantities: 1) the error in the PM of the
QSO, which is the quadrature sum of the centroiding errors from each
epoch, and 2) the error in the average PM (itself zero) of the
star field. This latter quantity gives us an idea of our systematic
errors by describing how accurately we were able to align the star
fields between the two epochs. In order to estimate it, the PM of each
star was determined as the difference between its average position in
epoch 2 and epoch 1, and then the error in the average PM of all the
stars was calculated.

\subsection{Consistency Checks}
\begin{figure}
\plotone{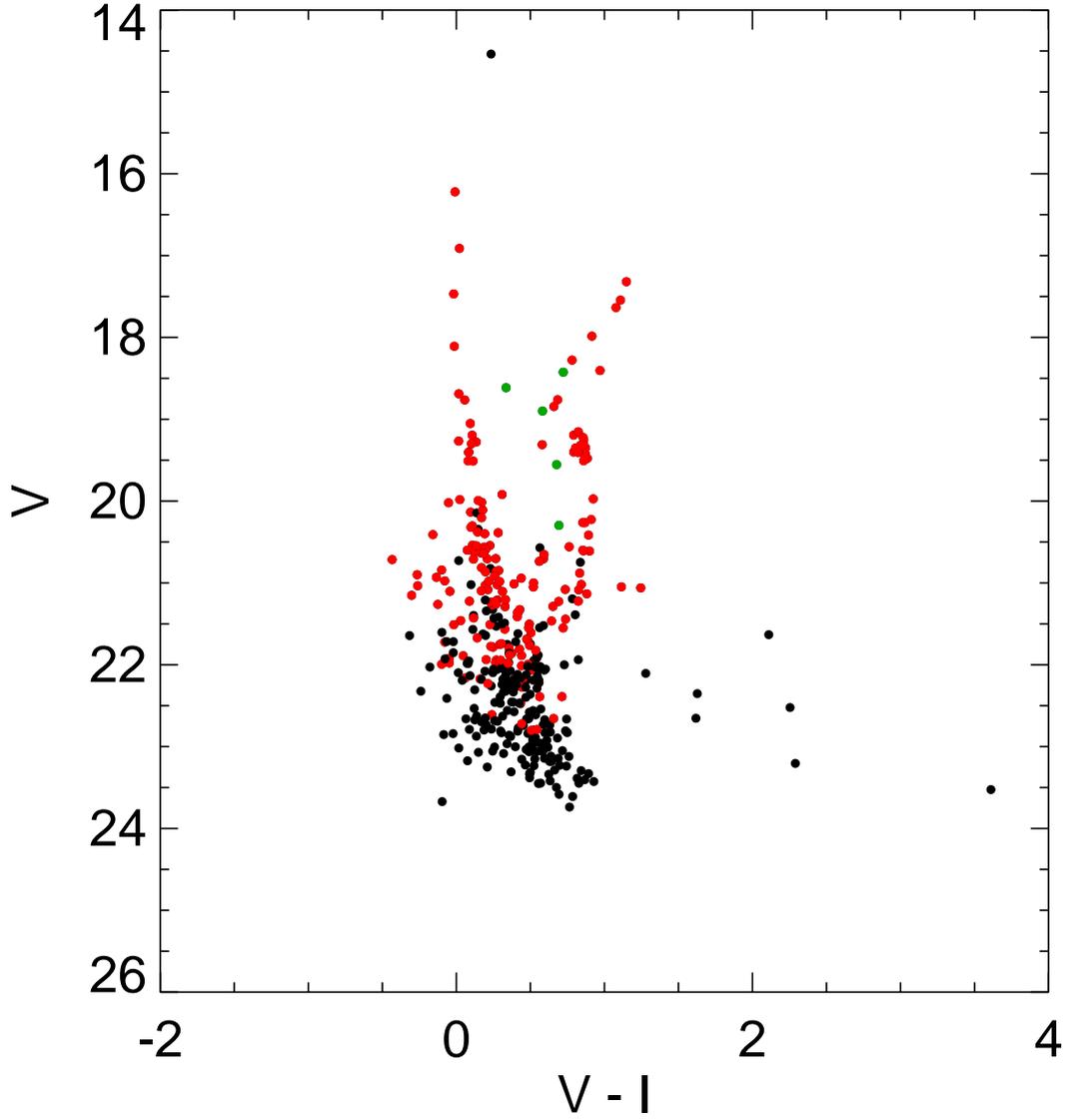}
\caption{ ($V-I$, $V$) color-magnitude diagram for the SMC. 
QSOs are marked in green, stars in the master-list with PM \& $\delta$PM
 $<0.1$ pixels are 
marked in red and the rest are shown in black.}
\label{CMD}
\end{figure}

Figure~\ref{CMD} is a ($V-I$, $V$) CMD of our SMC fields. QSOs are
shown in green, stars with PM and $\delta \rm{PM} < 0.1$ pixels (the
cut that we applied in PM and $\delta$PM space) are shown in red, and
the remaining sources in the master-lists are shown in black. The CMD
has the expected shape, with a clearly defined main sequence and giant
branch. There are a small number of red clump stars at $\sim V = 19.5,
\ V-I = 0.9$ (see Dolphin \etal 2001). This is, as expected,
approximately half a magnitude fainter than what we found for the LMC
and provides an additional calibration check. The QSOs appear spread
out over the CMD as we saw in the case of the LMC. As in the LMC
dataset, the QSOs behind the SMC are not systematically the brightest
sources in the images nor do they all have the same color. This
minimizes potential magnitude-based and color-based errors.

\begin{figure}
\plotone{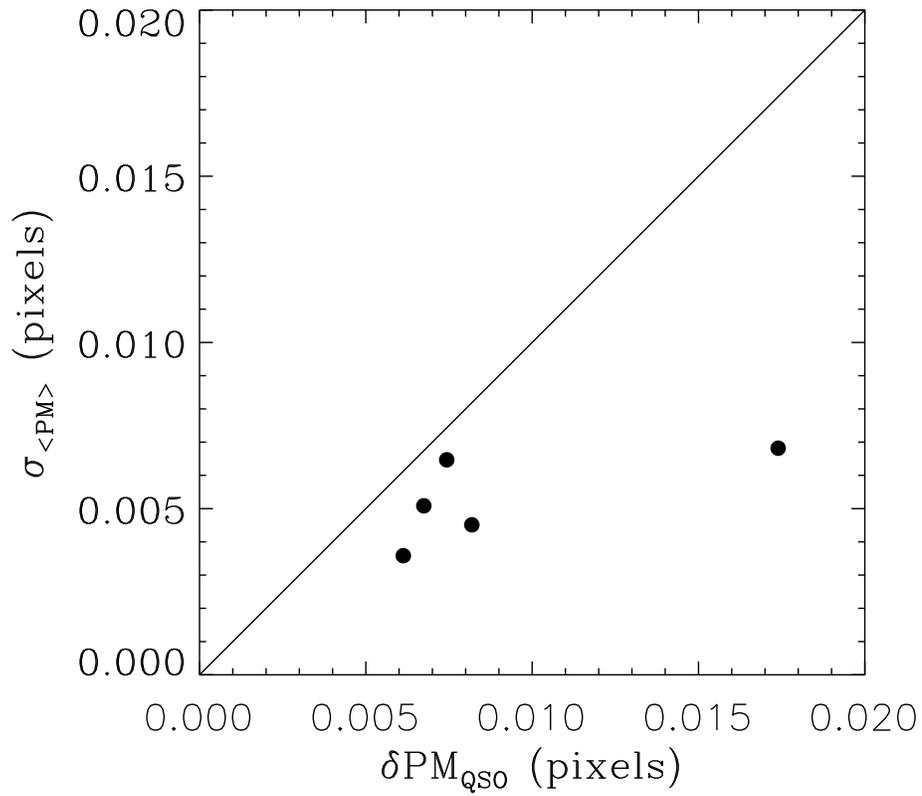}
\caption{The distribution of errors in our linear transformations for each 
field. The $x$-axis shows the error in the PM of the QSO and the 
$y$-axis shows the error in the average PM of the star-field.}
\label{distoferrors}
\end{figure}

Figure~\ref{distoferrors} shows the final QSO PM errors ($\delta
\rm{PM_{QSO}}$) versus the final error in the average motion of the
star-field between epoch 1 and epoch 2 ($\sigma_{\rm{<PM>}}$). The
average motion of the star-field itself was set up to be zero in our
linear transformations and any deviation from zero is a signal of
systematic errors. The plot shows that the final errors are dominated
by the centroiding errors of the QSOs and that aligning the star-field
introduces smaller errors in comparison. This is what we expected
given that many stars were used to fit for the terms of the linear
transformations. 

\begin{figure}
\includegraphics[width=.8\textwidth]{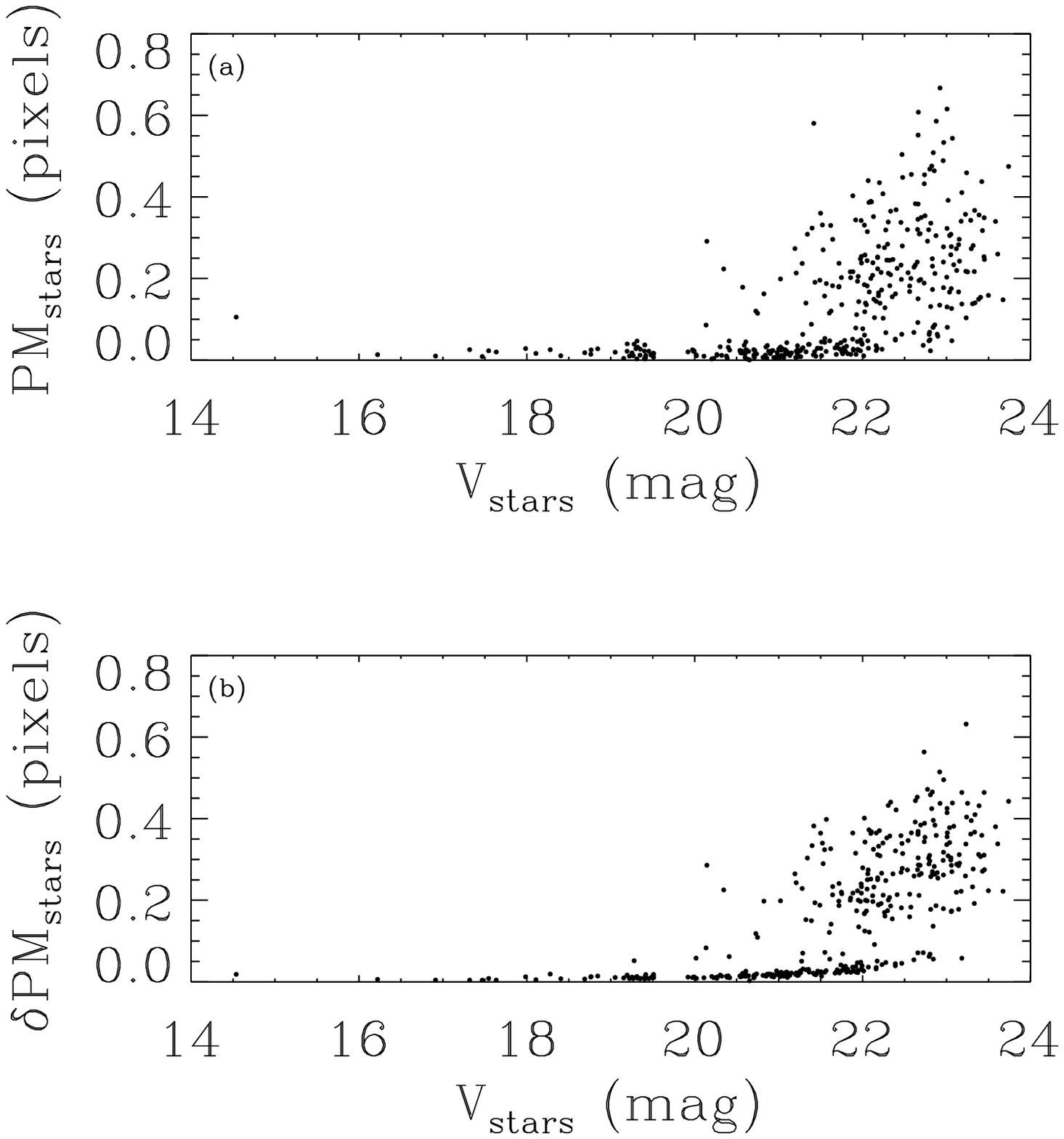}
\includegraphics[width=.8\textwidth]{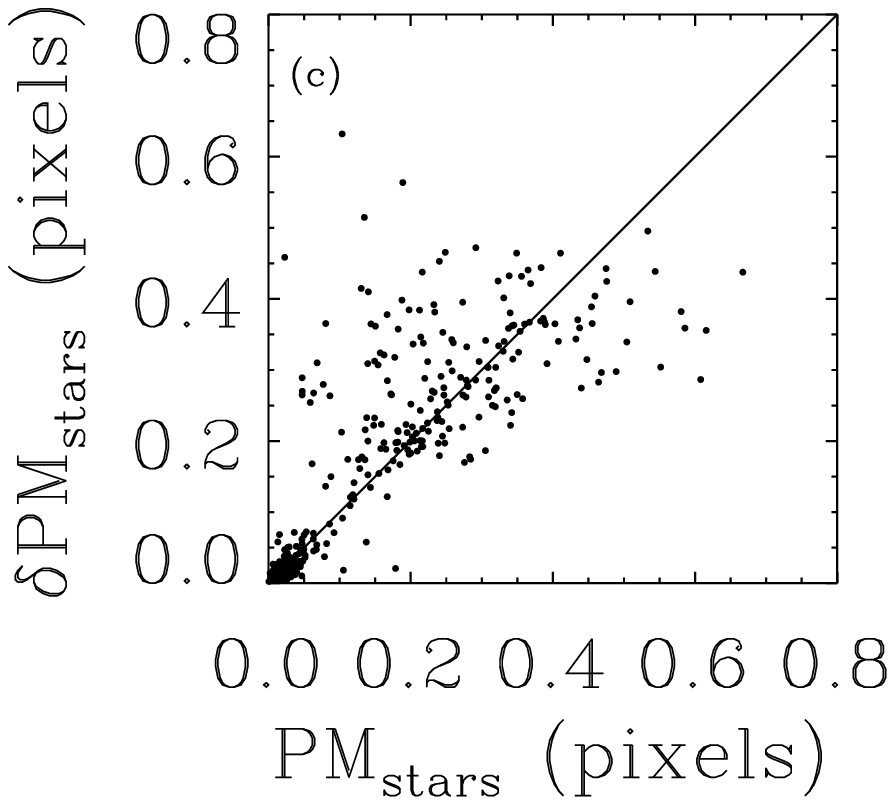}
\caption{(a) PMs for all stars in all master-lists 
as a function of their $V$ magnitude; (b) $\delta$PM for all stars in all 
master-lists as a function of $V$ magnitude; (c) $\delta$PM vs. PM for all 
stars in all master-lists.}
\label{errorcheck}
\end{figure}

\begin{figure}
\plotone{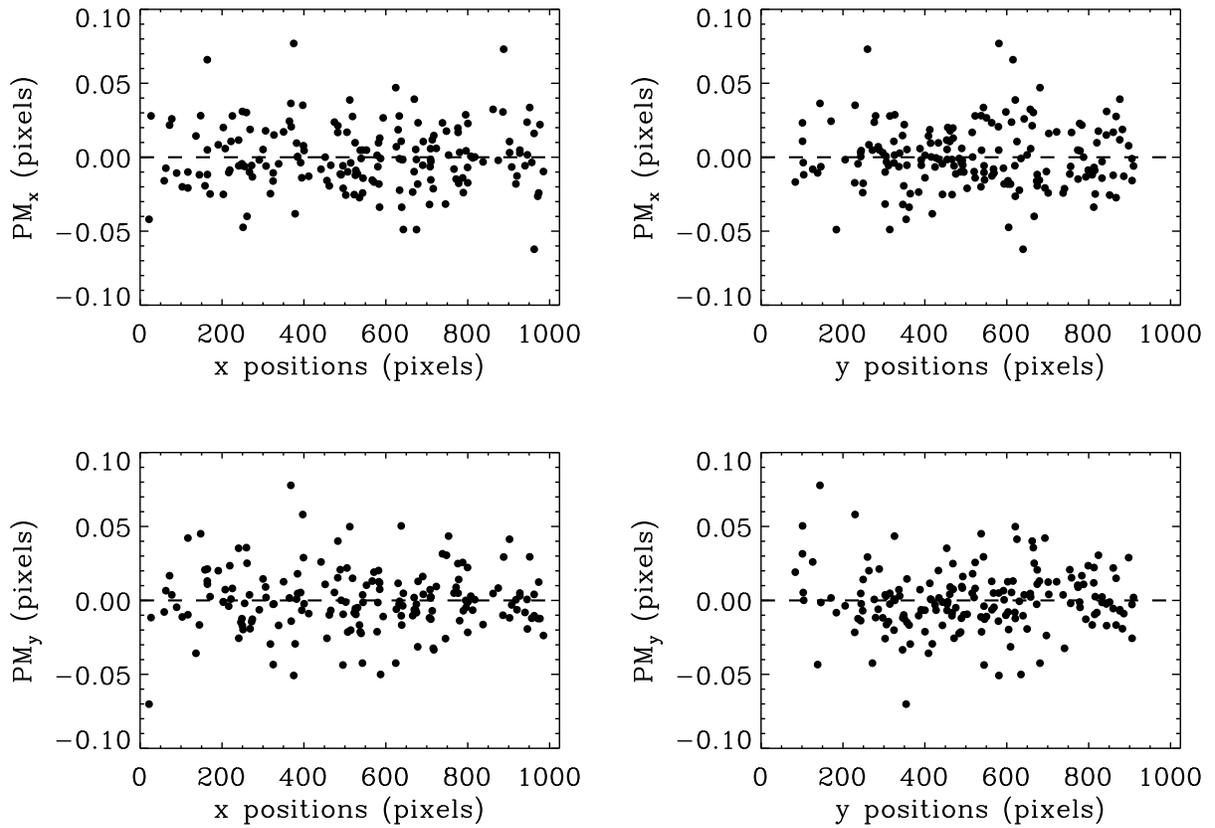}
\caption{PMs of 
the stars in the masterlist that have PM and $\delta \rm{PM} <0.1$ pixels 
versus chip location separately for $x$ and $y$ to 
see if there are any systematic trends with position.}
\label{pmpos}
\end{figure}

\begin{figure}
\plotone{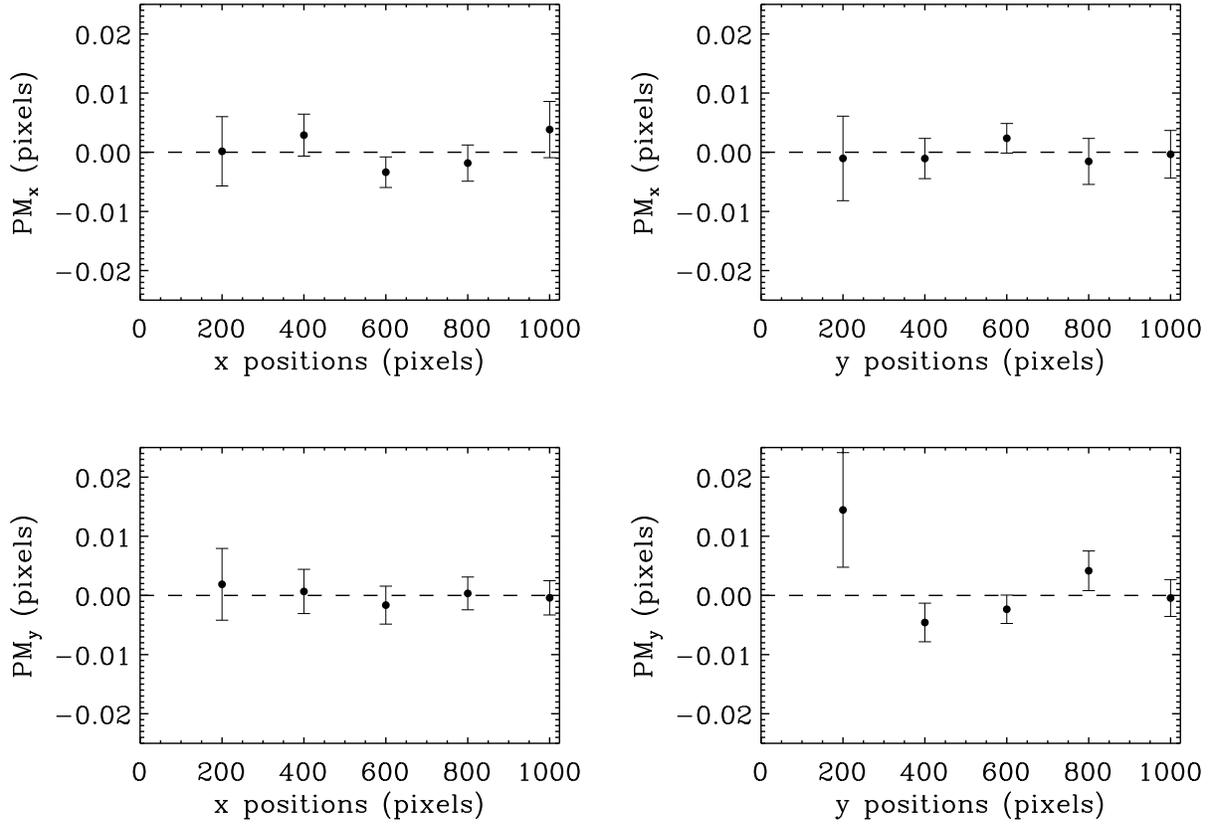}
\caption{Average PMs of 
the stars in the masterlist that have PM and $\delta \rm{PM} <0.1$ pixels 
versus chip location. The PMs of the stars have been binned for every 200 
pixels and then averaged.}
\label{pmbins}
\end{figure}

Figures~\ref{errorcheck}(a) \& (b) show the PMs and $\delta$PMs of all
the stars in the master-lists as a function of $V$ magnitude.  The
plots have the expected $(S/N)^{-1}$ shape. The apparent discontinuity
at $\sim0.1$ pixels is a result of the cuts applied in the stars that
were used to solve for the linear
transformations. Figure~\ref{errorcheck}(c) shows $\delta$PM versus PM
for all the stars in the master-lists. It shows that the PM residuals
for the stars are not much larger than the expected random errors. The
next two figures seek to address the possibility of any systematic
trends as a function of chip location. Figure~\ref{pmpos} is a plot of
the PMs of the stars which have PM \& $\delta \rm{PM} < 0.1$ pixels
versus $x$ and $y$-position on the chip separately. There is no
obvious trend with position on the chip and the scatter looks
comparable in both $x$ and $y$.  Figure~\ref{pmbins} is a plot of the
average PM value for every 200 binned pixels in an effort to get down
to the level of the noise. As can be seen from this plot, there is no
evidence for systematic errors larger than $\sim 0.005$ pixels, which
confirms the findings of AK04.

\section{Results}
\subsection{Proper Motion Results for the SMC}
Table 2 presents the results for each QSO field obtained using the
procedure discussed in \S~3.1. $N_{\rm{sources}}$ refers to the 
number of real sources in each field (present in at least half of the 
frames), $N_{\rm{master}}$ refers to the 
number of stars in the master-list, and $N_{\rm{used}}$ is the 
number of stars  used in the final transformation after the 
iteration scheme had converged.  Each observed QSO PM provides an
independent estimate of the PM of the SMC center of mass. Having
obtained these estimates, we are now in a position to check for any
further systematic trends on a per-field basis.

\begin{figure}
\centerline{
\epsfxsize=0.75\hsize
\epsfbox{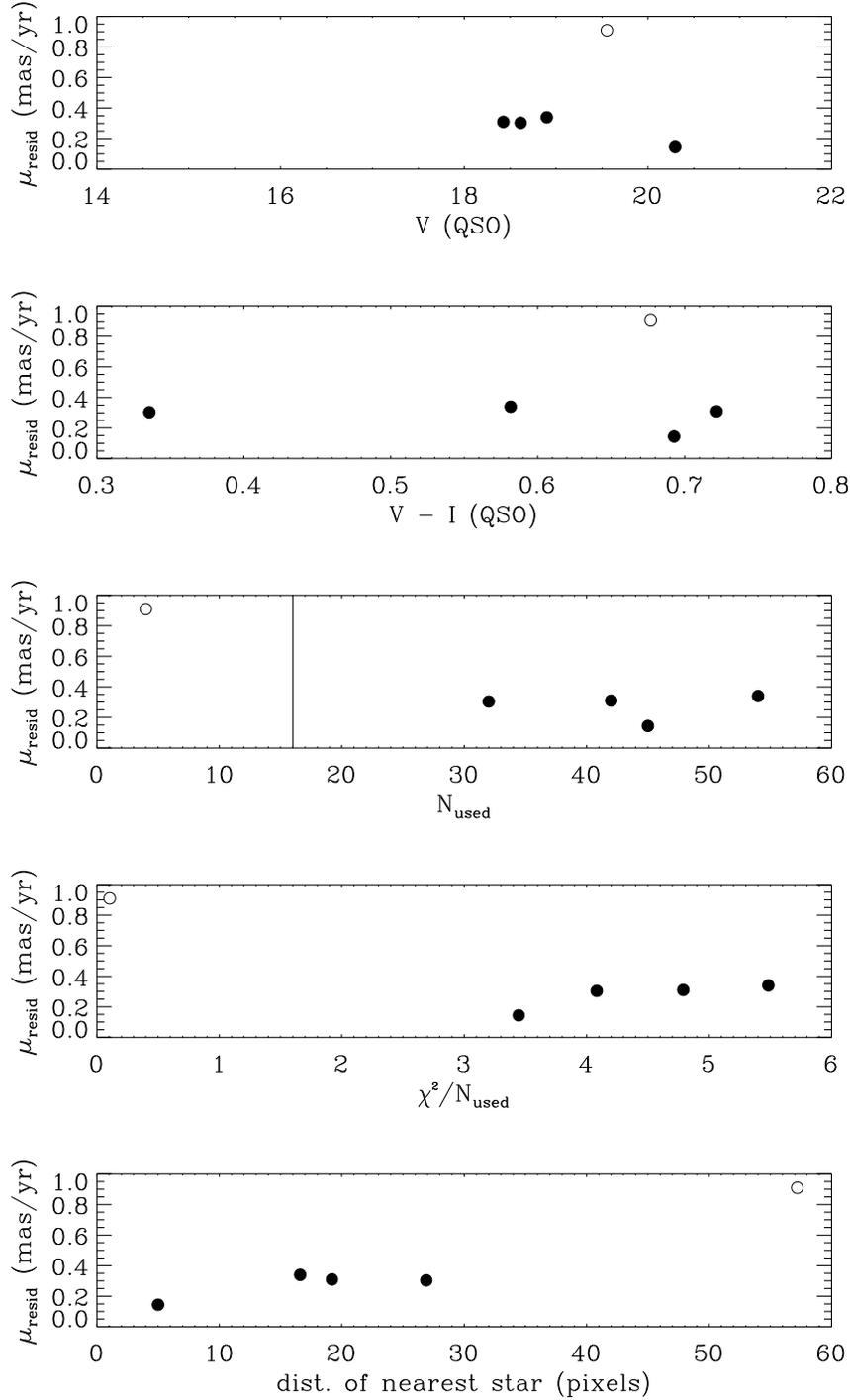}}
\caption{Plots of $\mu_{\rm {resid}}$ as a function of QSO $V$
magnitude, $V-I$ color, $N_{\rm {used}}$, $\chi^2/N_{\rm {used}}$ and
distance to the nearest neighboring star. We used the same criteria
here as in Paper~I, retaining only those fields with $N_{\rm {used}} >
16$ and $\chi^2/N_{\rm {used}} < 15$ (vertical lines) in our final
sample (closed circles). The field that is rejected on the basis of
these cuts (S4) is shown with an open circle in each panel.}
\label{finalcuts}
\end{figure}

Figure~\ref{finalcuts} is a plot of the residual PM ($\mu_{\rm
resid}$) for each QSO field as a function of $V$ magnitude, $V-I$
color, ${N_{\rm used}}$, $\chi^2/N_{\rm{used}}$ and distance of the
QSO from its nearest neighbor star. The residual PM is measured with
respect to our final PM estimate for the SMC (see
equation~(\ref{PMfinal})).  An analogous plot in Paper~1 alerted us to
systematic effects as a function of ${N_{\rm used}}$ and
$\chi^2/N_{\rm{used}}$, and consequently, we made cuts in the LMC
dataset which amounted to using only those fields with ${N_{\rm
used}}>16$ and $\chi^2/N_{\rm{used}}<15$ . We have made the same cuts
in the current dataset. This eliminated one field, S4, that has a very
low value of ${N_{\rm used}}$.  S4 is a particularly sparse field
located towards the North-West of the SMC (see
Figure~\ref{figure1}). It has a very discrepant PM (see Table~2) and
is shown with an open circle in Figure~\ref{finalcuts}. The remaining
fields are shown with closed circles. The PM estimates for these
remaining fields show no obvious systematic trends associated with any
of the above quantities.

\begin{figure}
\plotone{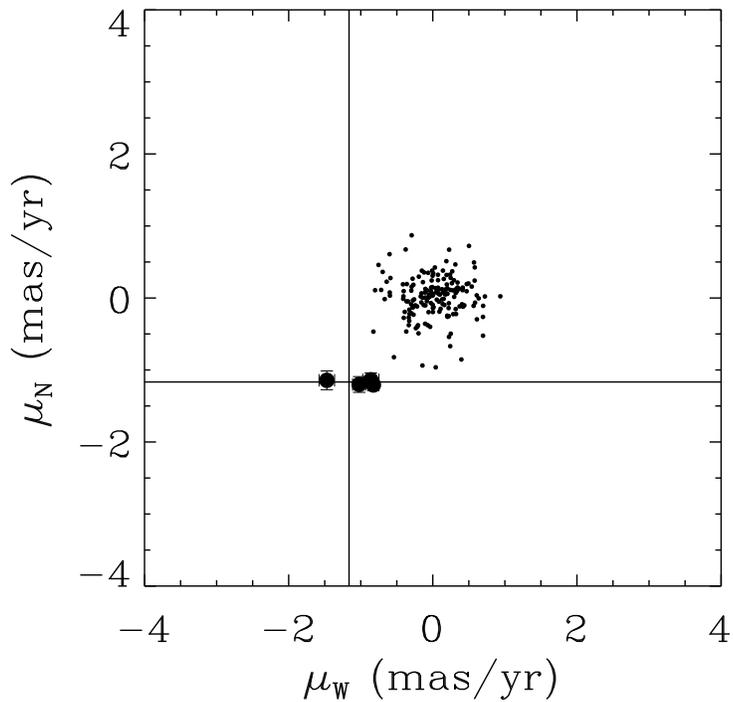}
\caption{The observed PM ($\mu_{W}$, $\mu_N$) for the 4 QSO fields
(i.e., $-1 \ \times$ the observed reflex motion of the QSO; columns 5
\& 6 of Table~2) that pass all our criteria. The residual PMs of the
SMC stars in the fields are plotted with open circles. The reflex
motions of the QSOs clearly separate from the star motions.  The
straight lines mark the weighted average of the 4 fields, as listed in
equation~(\ref{PMfinal}).}
\label{PMs}
\end{figure}

Figure~\ref{PMs} is a plot of the QSO PMs for the 4 remaining fields
in the $(\mu_W, \mu_N)$-plane. They are shown in comparison to the
proper motions of the SMC stars used in the analysis. The
stars are shown with points and the QSOs with filled circles. The PMs
of the stars cluster around zero as expected given their PM errors
(see Figure~\ref{errorcheck} (a) \& (b); the error bars are not shown
here). The reflex motions of the QSOs are clearly distinguishable
from the star motions. The solid lines mark our final value for the
SMC PM presented in equation~(\ref{PMfinal}) and discussed below.

The four available measurements for the SMC PM agree very well, to
within $\pm 0.04 \masyr$, in the North-South direction (see
Table~2). However, in the East-West direction there is agreement only
to $\pm 0.32 \masyr$, indicating the presence of unidentified
systematic errors. This is not surprising in view of our results
obtained for the LMC in the context of the same observing program. The
RMS scatter of the proper motion measurements from each LMC field was
larger than the scatter expected from random errors alone, indicating
the presence of systematic errors in our LMC measurement as well.

For the LMC, after using the same rejection criteria in terms of
${N_{\rm used}}$ and $\chi^2/N_{\rm{used}}$ as we do here we were left
with 13 measurements. We estimated the final proper motion as the
weighted average of the $N=13$ measurements. The error in the result
was estimated as the RMS scatter divided by $\sqrt{N}$, as expected
for an ensemble of independent and uncorrelated measurements. This was
appropriate because the following three criteria were met: (1) the
measurements were obtained for different fields, at different
positions in the galaxy; (2) the observations were obtained at
different times of the year, with the relative orientation between the
North direction and the detector $y$-axis distributed more or less
randomly; and (3) the scatter in the final proper motion residuals
appeared to be distributed randomly.

The situation for the SMC is somewhat more complicated than for the
LMC, because neither the second nor the third criteria in the
previous paragraph appears to be met. Observations S1, S2 and S3 were
obtained with very similar telescope orientation (to within $\pm
6^{\circ}$) in both epochs, while observation S5 used a very different
orientation in the first epoch (see Table~1). Moreover, the proper
motion results for S1, S2 and S3 are in excellent agreement with each
other (to better than $\pm 0.1 \masyr$ in both coordinates), while the
result for S5 differs by $0.4$--$0.6\masyr$. So it appears that the
unknown systematic errors in the study are correlated with telescope
orientation. This is not surprising because any uncertainties in the
geometric distortion correction would have fixed orientation in the
detector frame.

The proper motion results suggest that it would be a mistake to treat
S1, S2, S3 and S5 as independent and uncorrelated measurements. That
would underestimate the error bars on the final proper motion
result. Instead, we first take the weighted average proper motion of
S1, S2 and S3. This gives $(\mu_W,\mu_N) = (-0.89 \pm 0.05, -1.18
\pm 0.05 \masyr$), where the listed errors account only for the propagation
of random errors. We denote this measurement as S123, and treat this
as one of only two independent measurements of the SMC proper motion
(the other one being S5). This is conservative in that it increases
the error bars on the final proper motion measurement. 

In our LMC proper motion study we found that the scatter between the
PM estimates obtained from different fields was larger than expected
based on the random errors alone. This indicates that the true error
for each field has not only a random component, but also a systematic
component. The observed scatter implies a systematic error of
0.24$\masyr$ per coordinate per field. For the LMC this error could be
estimated directly, because the large number of available fields
allowed a reliable calculation of the scatter. This is not true for
the SMC, because S123 and S5 provide only two independent
measurements. However, the SMC and LMC data were obtained with the
same setup, in the same observing program, and at the same general
time. So it is reasonable to assume that the systematic errors for the
LMC and SMC fields are the same. For both the S123 and S5 measurements
we then calculate a total error by adding in quadrature a systematic
error estimate of $0.24 \masyr$ per coordinate (based on the LMC
results) to the random errors. We then take the weighted average of
S123 and S5 as our final estimate for the SMC proper motion. This
yields
\begin{equation}
\label{PMfinal}
  \mu_W = -1.16 \pm 0.18 \masyr, \ \mu_N = -1.17 \pm 0.18 \masyr \quad 
  (HST).
\end{equation}
where the errors now take into account both random and systematic
errors. The final errors per coordinate are larger by a factor $2.8$
than what we obtained in our LMC proper motion study. This is
plausible, since one would naively expect the errors to scale as the
inverse square root of the number of measurements; and for comparison
$\sqrt{13/2} = 2.5$.

\begin{figure}
\plotone{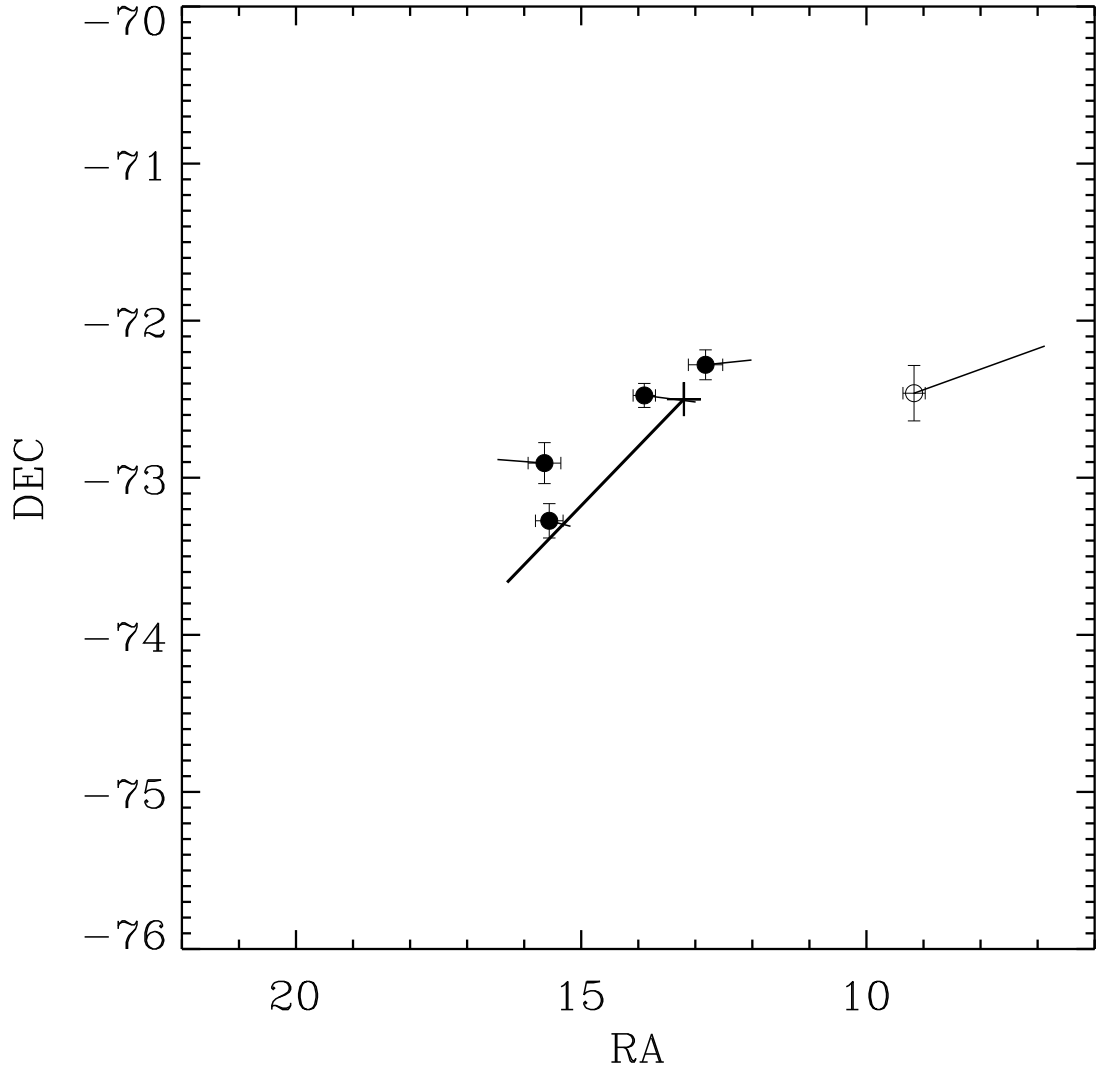}
\caption{Circles show the positions of the QSO fields. The field that
is rejected (from our final SMC PM estimate) is shown with an open
circle and the 4 remaining fields are shown with filled circles. The
error bars for each field are plotted as well. The vectors at these
circles show the residuals between the PM estimates derived from the
data for these fields and the adopted weighted average. The latter is
given in equation~(\ref{PMfinal}) and is shown by the bold solid
vector that is anchored by a plus sign.}
\label{residuals}
\end{figure}

Figure~\ref{residuals} is a vector plot of the residuals between the
PM estimates for each field and the adopted average
(equation~(\ref{PMfinal})), shown with the $1 \sigma$ error bars for
each field.  Circles show the positions of the QSO fields in RA/DEC
space. Closed circles represent the final 4 fields that were used in
our final estimate for the PM and the open circle shows the rejected
field S4. The thick vector anchored by a plus sign shows the size of
the inferred center of mass proper motion of the SMC at the adopted
SMC center. The vectors appear to be randomly oriented in the sky. In
particular, there is no evidence for any residual rotation of the
SMC. We discuss the rotation of the SMC in more detail in \S~4.2 below.

\begin{figure}
\plotone{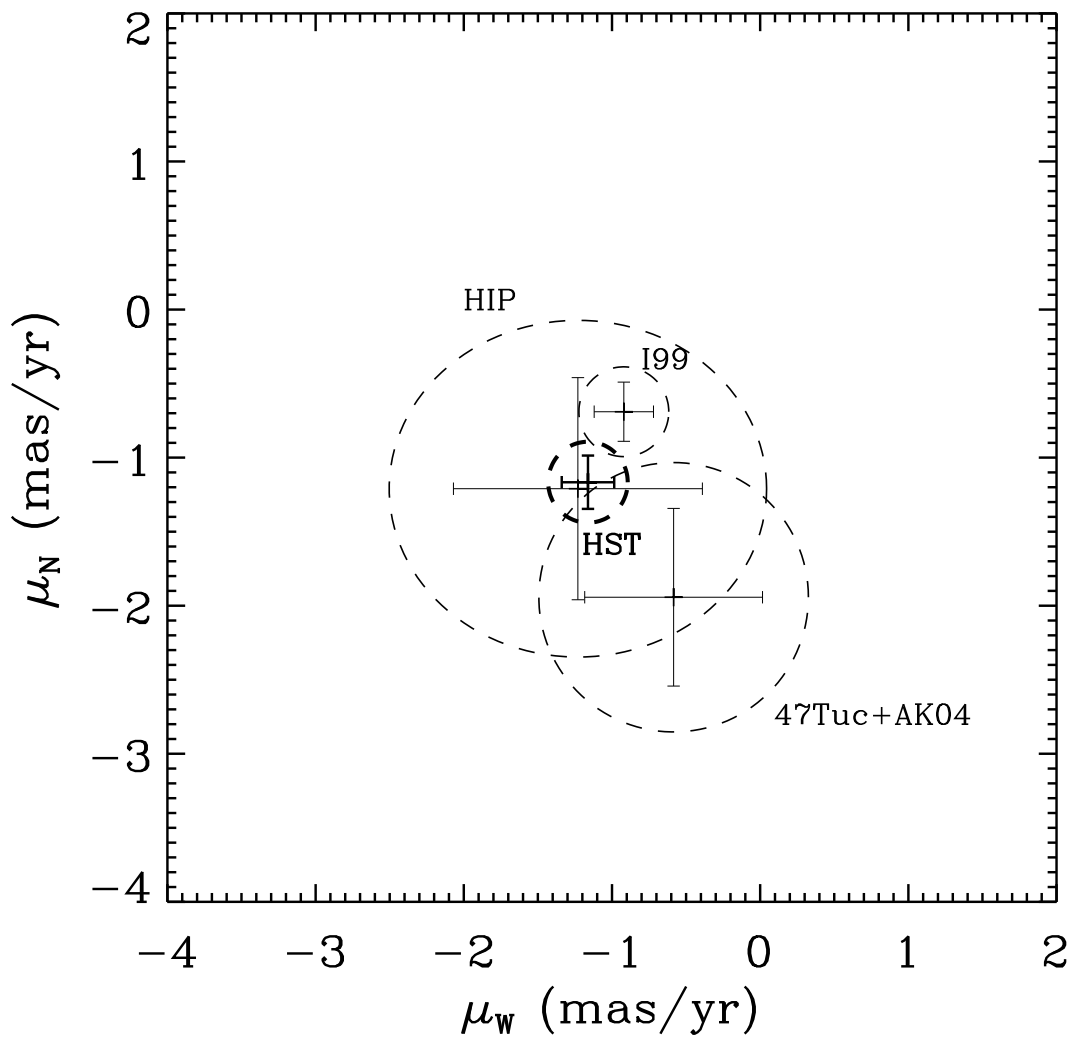}
\caption{Plot of the $(\mu_W, \mu_N)$-plane spanned by the proper
  motion of the SMC from various studies. Dashed ellipses indicate the
  corresponding 68.3\% confidence regions. The label HIP stands for
  the Kroupa \& Bastian (1997) Hipparcos study, I99 for the measurement 
quoted in Irwin (1999), 47Tuc$+$AK04 for the value obtained by combining the
  Freire \etal (2003) and AK04 studies, and HST for this study.}
\label{otherSMCpms}
\end{figure}

Our measurement of the SMC proper motion agrees to within $1\sigma$
with the one obtained in \S~1 as the unweighted average of the three 
existing measurements, $\langle \mu_W \rangle =-0.91 \pm 0.19 \masyr,
\langle \mu_N \rangle = -1.28 \pm 0.36\masyr$. However, our result has
either smaller errors or better understood errors than all previous
work.  The weighted average of our result with the quoted $\langle
\mu_W \rangle, \langle \mu_N \rangle$ is :
\begin{equation}
\label{PMusandthem}
  \mu_W = -1.04 \pm 0.13 \masyr, \ \mu_N = -1.19 \pm 0.16 \masyr \quad 
  {\rm (HST + other \ studies)} .
\end{equation}
We show our $HST$ measurement together with previous measurements in
Figure~\ref{otherSMCpms}, along with the corresponding 68.3\%
confidence ellipses. In the discussion that follows we adopt the
$HST$-only values given in equation~(\ref{PMfinal}).

\subsection{Three-Dimensional Space Motion of the SMC}
The proper motion of the SMC gives the following values for the $x$
and $y$ components of the SMC velocity in a Cartesian system on the sky,
with $x$ in the direction of west and $y$ in the direction of north:
\begin{equation}
(v_x, \  v_y) = (-340\pm52, -341\pm53) \kms,
\end{equation}
assuming a distance modulus of 18.95 (Cioni \etal 2000 and references
therein).  The corresponding transverse velocity, $v_t$, is $481 \kms$
at a position angle, $\Theta_t = 135\degr$. These can be combined with
the observed line-of-sight velocity, $146 \pm 0.6 \kms$ (Harris \&
Zaritsky 2006), and corrected for the reflex motion of the Sun, to
give the full three-dimensional space velocity in a Galactocentric
rest frame. The latter consists of a Cartesian coordinate system ($X$,
$Y$, $Z$) with the origin at the Galactic center, the $Z$-axis
pointing toward the Galactic north pole, the $X$-axis pointing in the
direction from the Sun to the Galactic center, and the $Y$-axis
pointing in the direction of the Sun's Galactic rotation. Following
the procedure outlined in \S9.2 of van der Marel \etal (2002) we get:
\begin{eqnarray}
\mathbf{v}_{\rm SMC} =  (-87 \pm 48, \ -247 \pm 42, \ 149 \pm 37) \kms 
\nonumber, \\
v_{\rm SMC} = 302 \pm 52 \kms \nonumber, \\       
v_{\rm SMC, rad} =  23 \pm 7 \kms \nonumber, \\
v_{\rm SMC, tan} = 301 \pm 52 \kms,
\end{eqnarray}
for the three-dimensional velocity of the SMC and its radial and
tangential components. The LMC and SMC space velocities are summarized
in Table~3.

We do not model any internal rotation contributions to the center of
mass motion of the SMC. This is because a spectroscopic analysis of
the radial velocities of 2046 red giants in the SMC by Harris \&
Zaritsky (2006) indicates that the SMC has no intrinsic
rotation. These authors find a small velocity gradient across the SMC
of $8.3 \kms \rm deg^{-1}$ and argue that the origin of such a
gradient need not be internal rotation but could instead be due to the
differential projection of the SMC's tangential velocity along
different lines of sight. They calculate that a tangential motion of
$\sim 500 \kms$, which is close to the value that we measure in this
work, will result in an apparent velocity gradient of $8.7 \kms \rm
deg^{-1}$ over the face of the SMC. They argue that even if the SMC
has some small intrinsic rotation, its value is much smaller than the
velocity dispersion that they measure ($\sigma = 27.5 \kms$), and thus
that the SMC is primarily supported by its velocity dispersion. By
contrast, the HI gas in the SMC does show clear rotation (Stanimirovic
et al. 2004). This presumably indicates that the HI gas resides in a
more disk-like distribution than the stars. Even if the stars in the
SMC were to rotate as fast as the gas (which appears ruled out by the
Harris \& Zaritsky 2006 observations), then this still would not
significantly affect our results. At the position of our QSO 
fields, the component of the HI rotation velocity field projected onto
the plane of the sky amounts to a proper motion of only $\sim 0.09
\masyr$. This is within the error bars of our final result. Moreover,
our result is obtained as an average of different fields. If there
were rotation of the SMC stars, then the rotation velocity vectors for
the different fields would not align on the sky. Hence, they would
partially cancel out when averaged.


\section{The Orbits of the Clouds around the Milky Way}
\subsection{The Fiducial Model}
We are now in a position to ask how the orbits of the two Clouds have
evolved over the past several giga-years (Gyr). Using the six position
and six velocity components of both Clouds as initial conditions we
can obtain a solution for the Clouds' movements in a prescribed dark
halo model. As a fiducial model we use a scheme that has been used
many times before in the literature and was originally formulated by
Murai \& Fujimoto (1980; see also Gardiner \etal 1994; Gardiner \&
Noguchi 1996; Bekki \& Chiba 2005), albeit in the absence of
tangential velocity information, especially for the SMC.

In order to derive the orbits of the Clouds, we first need the current
tangential velocities (from this study), line of sight velocities and
three-dimensional position parameters (from the literature). A summary
of the orbital parameters that we use is given in Table~3. In addition
to the current phase-space parameters, we need to assume models or
values for the following: the gravitational potential of the Galaxy,
the gravitational potential of the Clouds, the total masses and mass
profiles of the Clouds, the dynamical friction between the Clouds and
the Galactic halo, and that between the Clouds themselves. So there is
a large number of physical assumptions and parameterizations necessary
to calculate orbits. There has been much progress in the arena of
galaxy models (Hernquist 1990; Navarro, Frenk \& White 1997), and
there might be reason to deviate significantly from prior
approaches. However, the goal in the present paper is merely to see
what we obtain for the orbital evolution of the Clouds when we combine
our new observational data with a typical model that has been used
before. Thus we follow the basic scheme devised by Murai \& Fujimoto
(1980). Once we have characterized this conventional prescription, we
then ask in a forthcoming paper which, if any, of the parameters need
to be changed in order to better match observed features of the MC-MW
system such as the Magellanic Stream.

\noindent \textit{Gravitational Potentials:} \\
We represent the gravitational potential of the Galaxy, as a function
of distance from the center, by an isothermal halo distribution:
\begin{equation}
\phi_G(r) = -V_0^{2} \ \ln  r,
\end{equation}
where $V_0 = 220 \kms$ is the circular velocity which is assumed to be
constant far outside the disk. The total mass enclosed within a sphere
of radius $r$ kpc is
\begin{equation}
M_G(<r) = 5.6 \times 10^{11}\left(\frac{V_0}{220 \kms}\right)^2 
\left(\frac{r}{50 \rm kpc}\right)M_\sun.
\end{equation}
It is possible that the Galactic potential beyond the present location
of the Clouds is triaxial, and in principle, the results of this study
in combination with the morphology of the MS can be used to test
this. However, a significant departure from spherical symmetry would
induce a precession in the orbital plane of a satellite (e.g. Fellhauer
\etal 2006). The lack of a noticeable warp in the Stream in either
position in the sky or radial velocity space suggests that the
assumption of spherical symmetry is a good one at least to first order
(Lin \etal 1995), although admittedly we do not have any distance or
proper motion information for the Stream. We do not employ a cut-off
radius for the extent of the dark halo. This is a reasonable
assumption since the halo is thought to enclose the current orbits of
the Clouds.

The LMC and SMC are represented using Plummer models:
\begin{equation}
\phi_{L,S}(r) = GM_{L,S}/[(\mathbf{r} - \mathbf{r}_{L,S})^2 + K^2_{L,S}]^{1/2},
\end{equation}
with effective radii ($K_L, K_S$) of 3 and 2 kpc respectively.

\noindent \textit{The Masses of the Clouds:}\\ The most serious
uncertainty in the model input parameters arises from the lack of
precise determinations of the masses of the Clouds. The masses of the
Clouds will play a role in determining the amount of dynamical
friction that they feel from the MW, and they will determine to what
extent they are bound to each other. The latest observational data
imply a range of possible masses. A dynamical mass for the LMC within
8.9 kpc of $(8.7\pm4.3)\times10^{9}$ $M_\sun$ was obtained by van der
Marel \etal (2002) using an analysis of carbon stars. This is less
than half of the mass adopted in previous studies. For example,
Gardiner \& Noguchi (1996; hereafter GN96) adopted a value of $2\times
10^{10}$ $M_\sun$.  This was estimated by Schommer \etal (1992) from
radial velocities of several of the oldest star clusters in the LMC
that lie well beyond 6 kpc of its center. 

Our own measurements give a relative velocity between the Clouds at
  the current epoch of $105 \pm 42 \kms$. For comparison, we can
  calculate the escape velocity of the SMC from the LMC, $v_{e,\rm
  SMC}$, assuming a simple point mass geometry. Using the value of
  $2\times10^{10}$ $M_\sun$ for the mass of the LMC and $23$ kpc for
  the distance between the Clouds, we get $v_{e,\rm SMC} =
  \sqrt{2GM_L/r} = 87 \kms$. This is consistent with the observed
  relative velocity at $1\sigma$ confidence. But the LMC would need to
  have mass $M_L = 3 \times 10^{10} M_\sun$ if the Clouds are
  gravitationally bound and the relative velocity is as large as $105
  \kms$. Given the range of possible masses, we do not
  systematically search the parameter space of LMC mass in our models,
  but rather consider three cases for $M_{L} = 1, 2 \ \& \
  3\times10^{10}$ $M_\sun$.

For the SMC, a lower mass limit of $1\times10^{9}$ $M_\sun$ was
obtained by Hardy \etal (1989) from observations of carbon stars and
Dopita \etal (1985) from planetary nebulae. Both of these measurements
were made close to the SMC center, so its mass is probably much
larger. More recently, Zartisky \& Harris (2006) used a virial
analysis of the kinematics of 2046 red giant stars in the SMC
to obtain an enclosed mass within $1.6$ kpc of between $1.4$ and
$1.9\times 10^9 M_\sun$ and a mass within $3$ kpc of between $2.7$ and
$5.1\times10^9M_\sun$. This has prompted us to take a typical value
for the mass of the SMC of $M_S = 3\times10^{9}$ $M_\sun$. 

We do not include any effects of mass-loss in the fiducial
model. Mass-loss is probably significant given the amount of matter in
the MS (estimated to be $\sim2\times10^{8}$ $M_\sun$ according to
Putman \etal (2003)). This is obviously an oversimplification, but
should still serve as a basic picture of the overall dynamics of the
Clouds, given that we are not attempting to model the Magellanic Stream.

\noindent \textit{Dynamical friction:}\\
The effect of dynamical friction on the orbits of the Clouds as they
pass through the dark halo of the Galaxy is taken into account as in
previous studies using the expression from Binney \& Tremaine (1987):
\begin{equation}
{\rm F}_{G} = -0.428  \ln\Lambda_G \ \frac{GM_{L,S}^2}{r^2}.
\end{equation}
Here $r$ is the distance between each Cloud and the center of the
Galaxy. The Coulomb logarithm $\ln \Lambda_G$ of both Clouds is taken
as 3.0 (Binney \& Tremaine 1987). This assumes as a simplification
that the orbits are circular. Equation (7) is generally a good approximation
to the full expression for dynamical friction
\begin{equation}
\frac{d\mathbf{v}}{dt} = -\frac{4\pi \ln  \Lambda G^2 \rho M}{v^3}
\left\{[{\rm erf}(X) - \frac{2X}{\sqrt{\pi}}e^{-X^2}]
\mathbf{v}\right\}_{X = v/V_0},
\end{equation}
which does not make the assumption of circular orbits explicitly.  Our
experience shows that small differences only start to build up between
the two terms if the orbits are integrated quite far backward in time
(on the order of $\sim10$ Gyr).

We also take into account the dynamical friction between the LMC and
the SMC, following Bekki \& Chiba (2005): 
\begin{equation}
{\rm F}_{LS} = -0.428 \ln\Lambda_{LS} \frac{GM_S^2}{r^2_{LS}},
\end{equation}
where $\ln \Lambda_{LS} = 0.2$ and $r_{LS}$ is the distance between
the two Clouds. This force is assumed to act on the SMC when it comes
within the tidal radius of the LMC, which in this study is adopted as
15 kpc (van der Marel \etal 2002). The associated gain in angular
momentum to the orbit of the LMC is found to be negligible in our
models. This makes sense given the mass ratio between the two Clouds
($3 - 10$). Any energy gain felt by the LMC is not expected to affect
its orbital motion but would instead go into puffing up its halo
(e.g. Weinberg 2000). Thus we do not include any additional force
terms for the LMC. 

As we will discuss in \S~6.2, one of the striking results of our
models for the orbits of the Clouds is that it is very difficult to
keep them bound to each other for more than 1 Gyr in the
past. Dynamical friction between the Clouds, if significant, would
make this situation worse, since it would act to coalesce the orbit of
the SMC with that of the LMC and thus imply that they were on even
more disparate trajectories in the past. Thus for our fiducial model
we also investigate orbits in which we do not include dynamical
friction between the Clouds. This is justified as a first-order
solution because 1) the distribution of dark matter in the LMC is not
very well known, and 2) if dynamical friction was a significant factor,
and the Clouds have been a bound system for approximately a Hubble
time, then they probably would have merged already.

\subsection{The Search for Bound Orbits}
We propagate the orbits of both Clouds backward in time for 9 Gyr
using the fiducial model described above and a leapfrog integration
scheme outlined in Springel \etal (2001). The equations of motions for
the Clouds about a stationary Galaxy can be written down using the
equations in \S~5.1:
\begin{equation}
\frac{d^2\mathbf{r}_{L}}{dt^2} = \frac{\partial}{\partial\mathbf{r}_{L}}
[\phi_S(\mid\mathbf{r}_{L} - \mathbf{r}_{S}\mid) + 
\phi_G(\mid\mathbf{r}_{L}\mid)] + \frac{{\rm F}_{L}}{M_L}
\frac{\mathbf{v}}{v}
\end{equation}
and
\begin{equation}
\frac{d^2\mathbf{r}_{S}}{dt^2} = \frac{\partial}{\partial\mathbf{r}_{S}}
[\phi_L(\mid\mathbf{r}_{S} - \mathbf{r}_{L}\mid) + 
\phi_G(\mid\mathbf{r}_{S}\mid)] + \frac{{\rm F}_{S}}{M_S}
\frac{\mathbf{v}}{v}.
\end{equation}
For our mean estimates of proper motion (quoted in equation~(3) in
Paper~I and equation~(\ref{PMfinal}) in this paper), and ignoring the
error bars, we find that the Clouds become unbound from each other
very quickly in the past. It is therefore interesting to see if
\textit{any} orbits in our proper motion error ellipses will remain
bound for a significant fraction of a Hubble time.

We used a simple Monte Carlo scheme to draw twelve initial phase-space
coordinates from the errors subtended by the parameters in
Table~3. From these initial phase-space coordinates we then calculated
the initial values of position and velocity for both Clouds in a
Galactocentric reference system. We then propagated the orbits of both
Clouds backward in time for 9 Gyrs using the fiducial model described
above. This procedure was repeated 10,000 times.  For each orbit we
kept track of when the Clouds moved more than 50 kpc from each other
and labeled this time as the ``time of disruption'' of the bound
system (following GN96). Further, we repeated this exercise for a few
cases of Cloud mass, each with the inclusion of dynamical friction
between the Clouds and without.

\begin{figure}
\centerline{
\epsfxsize=0.75\hsize
\epsfbox{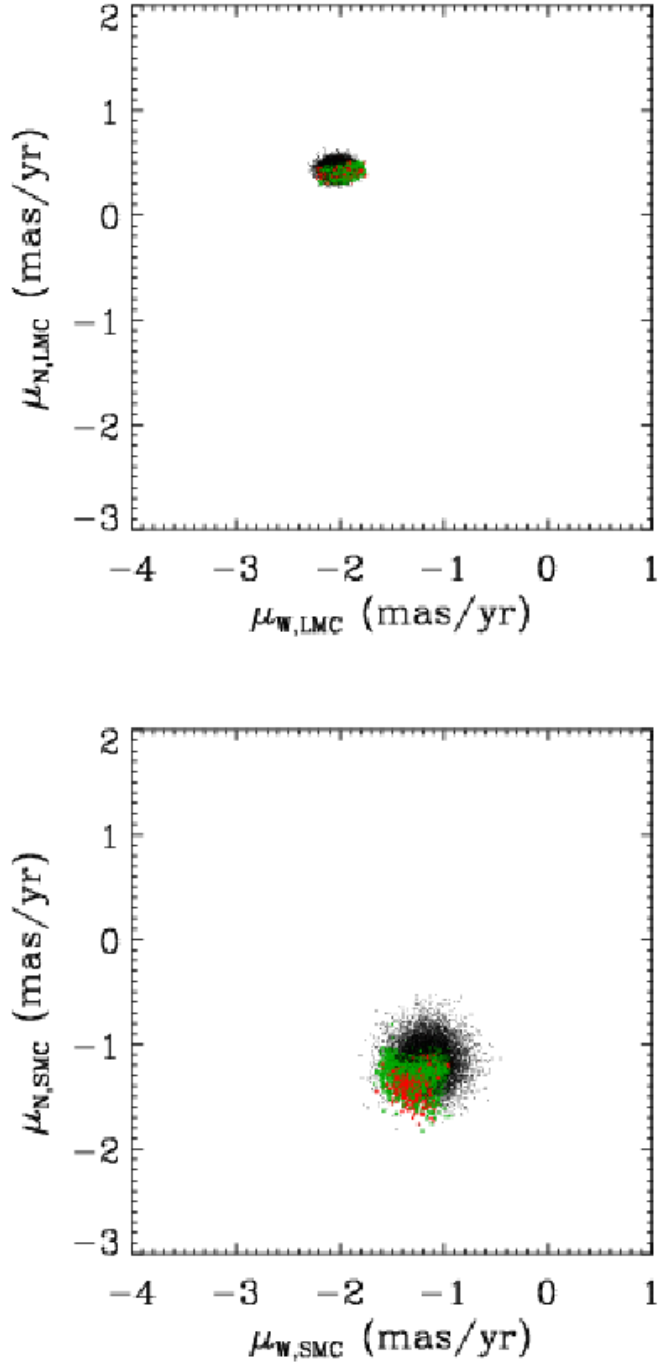}}
\caption{The past duration of the bound state of the Magellanic
  Clouds shown in the ($\mu_W, \ \mu_N$)-plane. The top panel shows
  the 10,000 initial proper motions drawn at random from the error
  ellipse of the LMC and the bottom panel shows the corresponding
  proper motions drawn from the error ellipse of the SMC. The duration
  of the bound state is represented by different colors, black for
  $<1$ Gyr, green for between 1 \& 5 Gyr and red for $>5$ Gyr.}
\label{MCresults_pm}
\end{figure}

\begin{figure}
\plotone{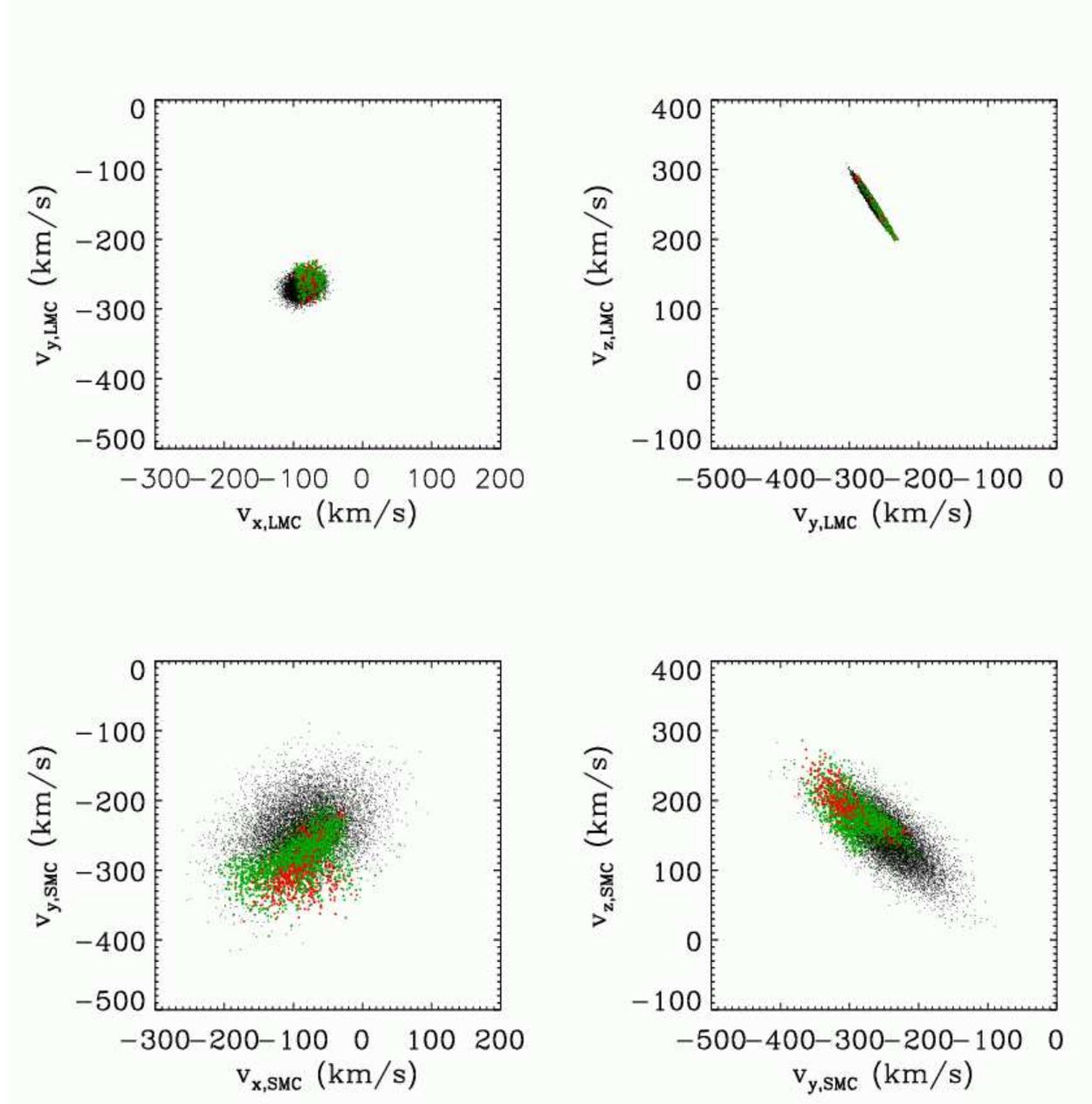}
\caption{The past duration of the bound state of the Magellanic
  Clouds shown in Galactocentric velocity space. The top panel shows
  the initial velocities calculated from the proper motion error
  ellipse of the LMC and the bottom panel shows the same for the
  SMC. The duration of the bound state is represented by different
  colors, black for $<1$ Gyr, green for between 1 \& 5 Gyr and red for
  $>5$ Gyr.}
\label{MCresults_v}
\end{figure}

For each orbit in a given Monte Carlo run, we kept track of the
initial proper motions (at time zero, i.e. the present time) and
color-coded them according to outcome : red if the Clouds stayed
together for more than 5 Gyr (``bound orbits''), green if they stayed
together for between 1 and 5 Gyrs, and black if they disrupted within
a Gyr.  Figures~\ref{MCresults_pm} and \ref{MCresults_v} are
representative of the results of the exercise as a whole but show the
specific case of $M_{L} = 3\times10^{10} \ M_\sun$ and $M_{S} =
3\times10^9 \ M_\sun$ without the inclusion of dynamical friction
between the Clouds.  Figure~\ref{MCresults_pm} shows the results in
proper motion space and Figure~\ref{MCresults_v} shows the results in
velocity space. Given our fiducial model, one can draw from a
significant fraction of the LMC error ellipse, although the southwest
portion is more favorable, and get orbits that have been bound for a
significant portion of the past Hubble time. For the SMC, bound orbits
are much more probable if we draw from the southeast portion of its
error ellipse. The locations of the mean values of the bound regions
are not overly dependent on the input masses of the Clouds, or the
inclusion of dynamical friction between the Clouds. In general,
however, a more massive LMC requires less of a shift in the proper
motions of both Clouds to bind the SMC, as expected. Thus we do find
bound orbits within our proper motion error ellipses.

\begin{figure}
\plotone{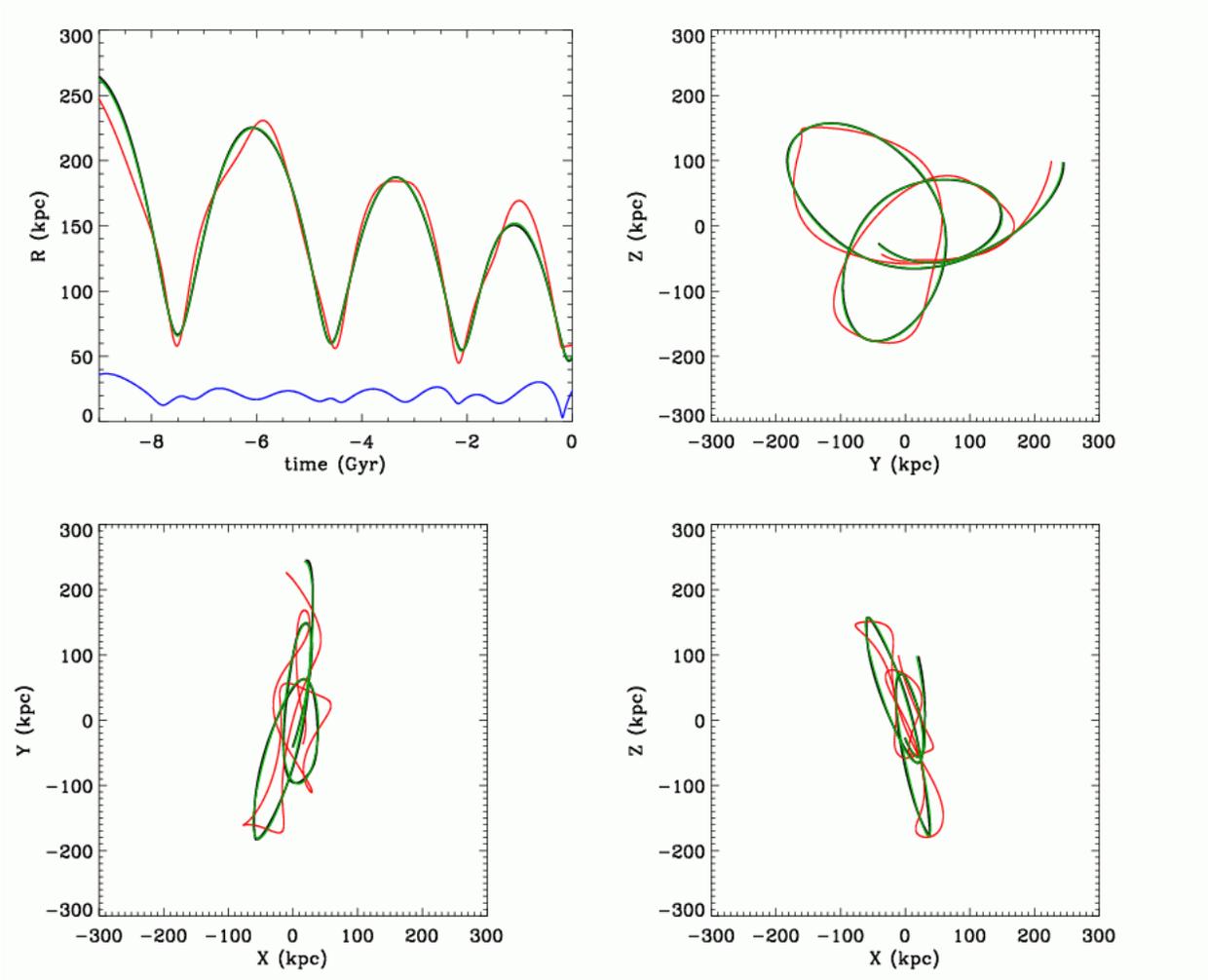}
\caption{A representative bound orbit from our simulations. Black
  shows the Galactocentric distance of the LMC, red shows
  Galactocentric distance for the SMC and green for the center of mass
  of the two Clouds. The blue line shows the distance between the
  Clouds.}
\label{orbit}
\end{figure}

Figure~\ref{orbit} is a representative bound orbit that we chose at
random from the simulation. The Galactocentric distance of the LMC is
shown in black, that of the SMC is shown in red and the Galactocentric
distance of the center of mass of the two Clouds is shown in green
(this is indistinguishable from the LMC's motion). The distance
between the Clouds is shown in blue. The center of mass has an orbital
period of $\sim2.5$ Gyr and an inclination of $103\degr$ (with respect
to the plane of the Galactic disk).

The number of bound orbits does depend on the masses of the Clouds and
whether dynamical friction between the Clouds is included. However,
this difference is not very significant. While the bound fraction
increases with increasing LMC mass, it is very small in all cases. The
bound fraction ranges from 3\% for LMC mass = $1\times10^{10}M_\sun$
to 10\% for LMC mass = $3\times10^{10}M_\sun$ for models without
dynamical friction between the Clouds included. It remains roughly at
2\% for models with dynamical friction between the Clouds. We discuss
the implications of our Monte Carlo simulations in the next
section. The fraction of bound orbits always increases when the LMC or
SMC masses are increased. On the basis of the presently available
kinematic data for these galaxies it cannot be ruled out that they
have massive dark halos that extend for tens of kpcs. Such halos would
make the galaxies more massive than has been assumed here, and would
increase the fraction of orbits consistent with our proper motion data
that are bound.

\subsection{Interpretation of Orbit Calculations}
The percentages of bound orbits in our Monte Carlo simulations (quoted
above) should not be interpreted as probability estimates for whether
the Clouds are bound or not. The small number of bound orbits is more
likely due to the large observational error bars and the comparatively small
phase-space spanned by bound orbits in a three-body problem. 
Our results are consistent with the outcome of most past searches for
bound orbits for the Clouds (see e.g. Appendix A of Bekki \& Chiba
2005, or Fig.~2 of Gardiner \etal 1994) in which authors have
found that a very small fraction of orbits searched in this
fiducial model will remain bound. Our SMC error bars are not small
enough to either confirm or rule out that the Clouds have been
a bound system. Given the results of our simulations, however, it is
worthwhile to further investigate whether there are indications that
some unidentified systematic error might be present in the data.

The proper motion of the LMC and the associated uncertainty should be
solid.  As demonstrated in Paper~I we had 21 QSO fields all taken at
different roll angles of the telescope, thus allowing any systematic
errors tied to the CCD frame to average out roughly as $1/\sqrt{N}$.
We did a number of consistency checks that showed no indication of
systematic effects. In the case of the SMC, however, we have far fewer
fields, and 3 out of the 4 reliable QSO fields were taken with the
same orientation of the camera. So if there were systematic errors
tied to the CCD frame (e.g. due to inaccuracies in the ACS HRC
geometric distortion correction) then they would not have averaged out
as they did for the LMC.  As a final consistency check on our data, we
have thus sought to answer the following question : if we keep the LMC
proper motion fixed at our best estimate values, what values of SMC
proper motion give rise to bound orbits? To answer this we kept the
model fixed at the fiducial model (with $M_L = 2\times10^{10}M_\sun, \
M_S = 3\times10^9M_\sun$), searched a grid in SMC proper motion space
from -2 to $2\masyr$ in the north-south and east-west directions, and
then looked for initial SMC proper motion values that gave rise to
bound orbits (for $>5$ Gyr).

\begin{figure}
\plotone{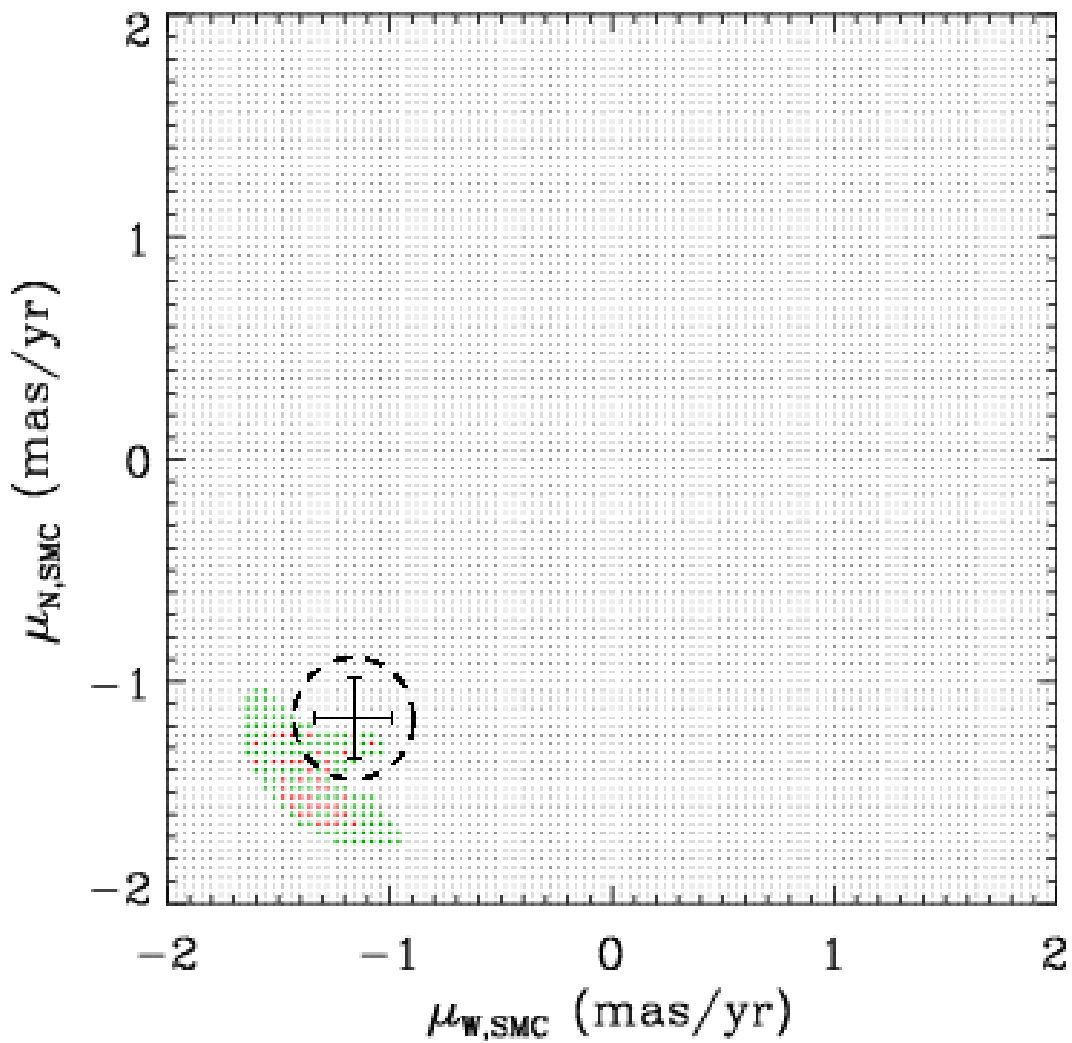}
\caption{The past duration of the bound state of the Magellanic
  Clouds as a function of the SMC's current proper motion. The length
  of the bound state is represented by different colors, black for
  $<1$ Gyr, green for between 1 \& 5 Gyr and red for $>5$ Gyr. The data
  point with error bars shows our measurement of the SMC's proper
  motion. The dashed ellipse is the corresponding 68.3\% confidence
  region.}
\label{SMCgrid_pm}
\end{figure}

\begin{figure}
\centerline{
\epsfxsize=0.7\hsize
\epsfbox{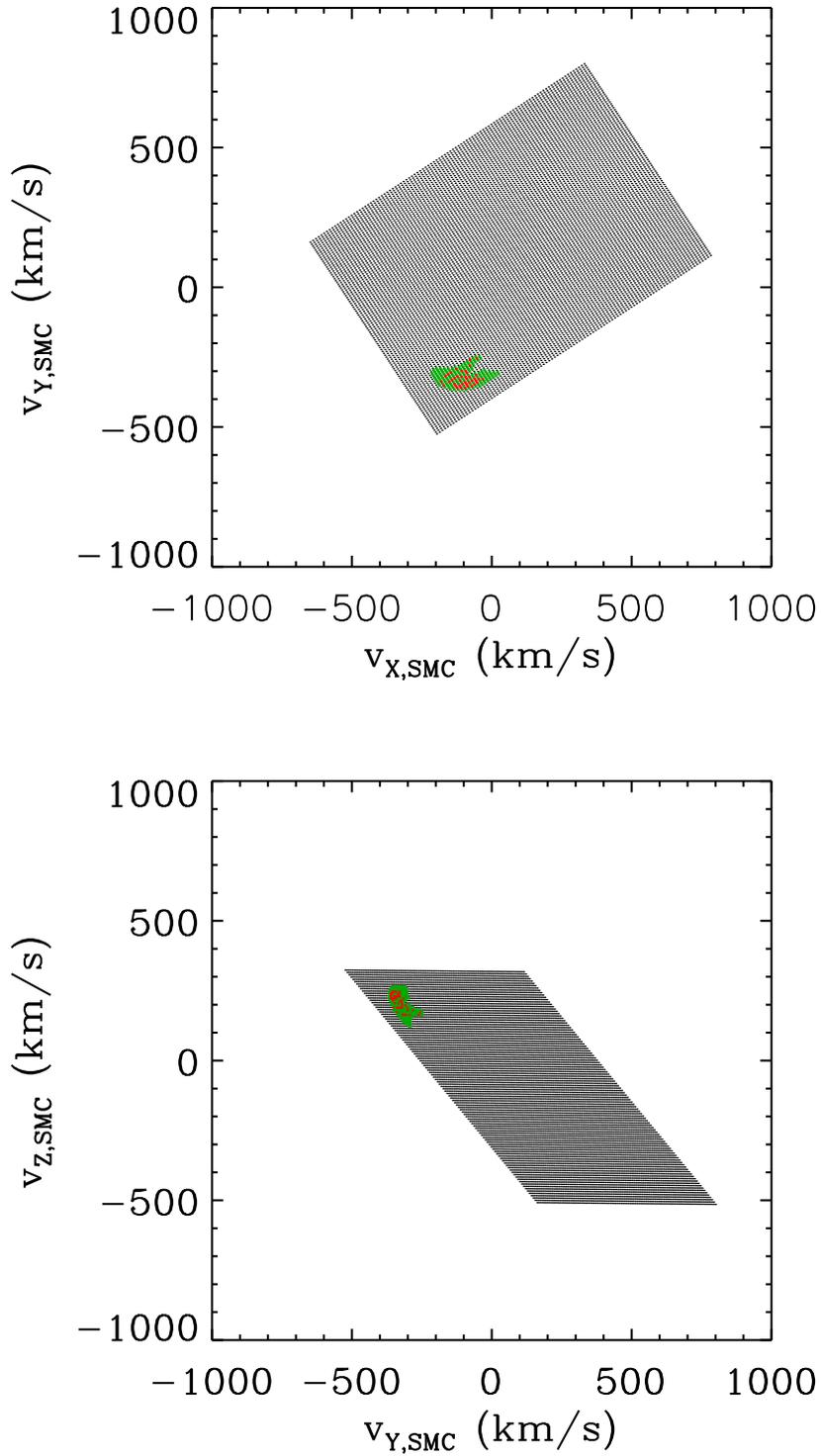}}
\caption{The past duration of the bound state of the Magellanic
  Clouds as a function of the SMC's current Galactocentric velocity.
  The length of the bound state is represented by different colors,
  black for $<1$ Gyr, green for between 1 \& 5 Gyr and red for $>5$
  Gyr.}
\label{SMCgrid_v}
\end{figure}

Figure~\ref{SMCgrid_pm} shows the results. It is a plot of the grid in
SMC proper motion space that was searched with the duration of the
bound state represented by points of different colors: black for $<1$
Gyr, green for between 1 \& 5 Gyr and red for $>5$
Gyr. Figure~\ref{SMCgrid_v} is the same but shows the results in
velocity space. The red region in Figure~\ref{SMCgrid_pm} has a mean
value of $\mu_W = -1.35 \pm 0.12 \masyr, \ \mu_N = -1.45 \pm
0.13\masyr$, which is an approximate $1\sigma$ shift from our measured
value of SMC proper motion in the westward direction, and
$\sim1.5\sigma$ in the northward direction. These values are stated
for reference only because the red region should not be interpreted as
a Gaussian error region. Every orbit in this region is acceptable, and
the only requirement is that the $1\sigma$ errors of our data (shown
with a black ellipse) have non-zero overlap with the red region. This
seems to be the case. What this means is that, given the error bars,
our measurement for the SMC proper motion falls where it is expected
to be based on the LMC proper motion and the assumption that the
Clouds are a bound system. Conversely, if both our LMC and SMC proper
motion measurements had some unidentified systematic error then this
would most likely have yielded orbits that are unbound. The fact that
bound orbits exist within the error regions of our data is therefore a
strong argument that our measurements and error estimates are
realistic.

\subsection{The `Recent  Coupling' Model}
Our SMC proper motion is consistent with a bound status for the
Clouds. But, while we do find bound orbits in our Monte Carlo
simulations, we also find many disrupted ones.  It thus remains
possible that the Clouds are not a bound system and have only
interacted long enough to produce the Stream. Most models of the
Stream suggest that this happened during their last perigalactic
passage. There is some evidence in the literature for the Clouds
having become dynamically coupled only recently. 

Bekki \& Chiba (2005) find that in their models, even with such small
total and relative velocities between the Clouds as those that have
gone into the theoretical modeling thus far (GN96), it is very hard
for the Clouds to maintain their bound status for very long backward
in time. They discuss a recent coupling scenario for the formation of
the Clouds, and argue that this has the following advantages. First,
it takes into account dynamical friction between the Clouds. Second,
the LMC has an asymmetric and irregular distribution of young clusters
and star-formation regions (van den Bergh 2000) and a scenario that
has the LMC uncoupled from the SMC might shed some light on their
formation history. Specifically, it might explain the `age gap'
problem in the LMC, i.e. the presence of only one Globular Cluster
(GC) which was formed between 13 \& 3 Gyr ago. Such a gap is not seen
in the SMC (Piatti \etal 2002, Da Costa 1991).  Since it is thought
that strong tidal perturbations trigger the formation of clusters
(e.g. Whitmore 1999), it is not clear why they would not have been
continuously produced in the LMC, since if the Clouds are a bound
system they would have been tidally perturbing each other for the past
15 Gyr. Bekki \& Chiba argue that this can be understood in terms of
the recent coupling model as follows: the LMC was formed as a low
surface brightness galaxy far enough from the Galaxy (they estimate
$150$ kpc from the old data) that the Galactic tidal field could not
trigger cluster formation till the LMC first encountered the SMC. The
SMC was formed closer and being less massive was more influenced by
the Galactic tide and formed GCs continuously. So to sum up, the
difference in cluster formation histories between the Clouds can
perhaps be explained by the differences in birth locations and initial
masses of the Clouds.

Despite the arguments of Bekki \& Chiba (2005), a model in which the
Clouds have always been a bound system is very compelling. The common
HI envelope that surrounds the Clouds indicates that they have been
associated with each other at least for some time. Given the sparse
distribution of outer satellites of the Galaxy, a capture event of the
SMC by the LMC seems improbable (GN96). The bound models pursued in
previous theoretical works have been able to reproduce the structure
and kinematics of the Magellanic Stream. Such models have also been
used to explain star formation about 0.2 Gyr ago in the LMC disk
(Gardiner \etal 1994), the structure of the stellar halo, and the
recent star formation history of the SMC (GN96; Yoshizawa \& Noguchi
2003). However, as Bekki \& Chiba (2005) point out, it is not clear
whether models that do not make the explicit assumption of bound
orbits could explain all these features equally well.

\section{Summary \& Future Work}
We obtained two epochs of ACS HRC data to determine the proper motions
of the LMC \& SMC with respect to a sample of background QSOs
distributed homogeneously behind the central few degrees of both
galaxies. Our result for the SMC is presented here. With 4 QSOs and an
approximately 2 year-long baseline, we have determined the proper
motion of the SMC to be $\mu_W = -1.16 \pm 0.18 \masyr, \ \mu_N =
-1.17 \pm 0.18 \masyr$ . This is the best available measurement of its
proper motion and is accurate to 15\%. We have carried out a suite of
tests to robustly quantify both random and systematic errors. We use
our LMC (from Paper~I) and SMC proper motion estimates to investigate
the past orbital evolution of both Clouds around the Milky Way. We
find that while our data are consistent with orbits in which the
Clouds have been bound to each other for approximately a Hubble time,
there are also many unbound orbits within the error circles. So even
though our errors on the SMC's motion are the most accurate thus far,
they are not sufficient to uniquely say whether the Clouds are indeed
bound. Also, it should be noted that our treatment of the orbits as a
restricted few-body interaction within a fixed Milky Way potential
is oversimplified. Structure formation in the Universe proceeds
hierarchically and it is unrealistic to assume that the mass and
properties of the galaxies involved were the same many Gyrs ago as
they are now. For example, over the past 9 Gyr the mass of the
Galaxy has probably increased by a factor of two (e.g. Bullock \&
Johnston 2005). Full cosmological simulations will be required to
take this into account properly.

Now that the space velocities of both Clouds are well-characterized,
it will be worthwhile to combine this information with the morphology
and radial velocity of the Magellanic Stream to place constraints on
the potential of the Galactic halo. The natural next step in this
project is therefore to vary the prescription of the Galaxy model and
quantify what type of halo would best match the Stream, given the
observed LMC \& SMC three-dimensional velocities. This will be the
subject of a forthcoming paper. The parameters that this should
constrain are the axial ratio of the halo, the slope of the circular
velocity curve beyond $\sim50$ kpc, and the implied Galactic mass at
that distance. There are few other kinematic tracers at this location
that can be used for such a measurement. Thus the combination of the
orbital information of the Clouds and the morphology of the Magellanic
Stream will provide a valuable new constraint. Combination of the
orbital information of the MCs and the Magellanic Stream with tidal
streams from other Local Group dwarfs such as the Sagittarius Stream
(Ibata \etal 1994; Majewski \etal 2003) might further constrain the
potential of the halo (e.g. Fellhauer \etal 2006; Belokurov \etal
2006; Johnston \etal 2005; Law \etal 2005; Helmi 2004).


The present-day velocities of the clouds assumed in models of the
Stream have spanned a considerable range. However, most authors who
have investigated tidal models for the origin of the Stream have used
the assumed present-day velocities suggested by GN96. These differ
from our observed values (given in Table~3) by $109 \pm 16\kms$ for
the LMC and $142 \pm 50\kms$ for SMC (where each listed value is the
length of the three-dimensional residual vector). So our measurements
are not consistent with these models. Therefore, even without detailed
calculations it is clear that revisions may be necessary to bring
existing models into accordance with our observations. Previous models
for the Stream have generally assumed a spherical logarithmic
potential for the dark halo. It might be necessary to deviate from
this simple prescription to obtain a good fit to the Stream given the
observed present-day velocities. As mentioned, this would provide new
insight into the properties of the Milky Way dark halo. But there are
many other uncertainties in existing models as well. Probably the
biggest uncertainties reside in the total mass and mass profiles of
the Clouds themselves. Future models may also need to incorporate more
details of the processes of hydrodynamics, mass loss, and star
formation. Conversely, tighter constraints on the dynamical evolution
of the Clouds may provide a better picture of the star formation
histories of the Clouds.

An important outcome of this work is the evidence that $HST$ is stable
enough to provide good proper motions using relatively short
baselines. However, future improvements to our estimate of the SMC's
proper motion should be possible by using a longer baseline and a
larger sample of QSOs. In addition to the Geha \etal (2003) MACHO
sample, QSOs behind the SMC have also been found from the OGLE
database (Dobrzycki \etal 2003). Even though the mass distribution of
the MCs remain the largest source of uncertainty, smaller errors on
the SMC's motion would greatly reduce the observational phase
space. This may allow us to definitively say whether the Clouds are
bound to each other.  Hence, an additional epoch of data would be
valuable.  With astrometric missions such as $SIM$ and $GAIA$ a rather
long time down the road, $HST$ can continue to make important
contributions to these subjects for several years to come.

\acknowledgments The authors would like to thank Jay Anderson for
making his geometric distortion calibration software available to
us. NK would like to thank Brant Robertson, Gurtina Besla, Lars
Hernquist and Matthew Holman for very useful discussions about
orbits. NK is also grateful to Dana Dinescu for pointing out some SMC
proper motion work in the literature, and to Pavlos Protopapas for
useful discussions about the project as a whole. Support for this work
was provided by NASA through grant numbers associated with projects
GO-09462 and GO-10130 from the Space Telescope Science Institute
(STScI), which is operated by the Association of Universities for
Research in Astronomy, Inc., under NASA contract NAS5-26555.

\section{References}
\noindent Anderson, J. \& King, I.~R.\ 2004a, ACS Instrument Science 
Report 04-15 (Baltimore: Space Telescope Science Institute)\\
\noindent Anderson, J., \& King, I.~R.\ 2004b, \aj, 128, 950\\ 
\noindent Belokurov, V., et al.\ 2006, \apjl, 642, L137\\ 
\noindent Bekki, K., \& Chiba, M.\ 2005, \mnras, 356, 680 \\
\noindent Binney, J.~ \& Tremaine, S.\ 1987, Galactic Dynamics,
(Princeton: Princeton Univ. Press)\\
\noindent Bullock, J.~S., \& 
Johnston, K.~V.\ 2005, \apj, 635, 931 \\
\noindent Caldwell, J.~A.~R., \& Coulson, I.~M.\ 1986, \mnras, 218, 223\\ 
\noindent Cioni, M.-R.~L., van der 
Marel, R.~P., Loup, C., \& Habing, H.~J.\ 2000, \aap, 359, 601\\
\noindent Connors, T.~W., Kawata D., Gibson B.~K.\ 2005
astro-ph/0508390\\
\noindent Cole, A.~A., Tolstoy, E., 
Gallagher, J.~S., \& Smecker-Hane, T.~A.\ 2005, \aj, 129, 1465 \\
\noindent Crowl, H.~H., Sarajedini, 
A., Piatti, A.~E., Geisler, D., Bica, E., Clari{\'a}, J.~J., \& Santos, 
J.~F.~C.\ 2001, \aj, 122, 220 \\
\noindent Da Costa, G.~S.\ 1991, IAU 
Symp.~148: The Magellanic Clouds, 148, 183 \\
\noindent Dobrzycki, A., Macri, 
L.~M., Stanek, K.~Z., \& Groot, P.~J.\ 2003, \aj, 125, 1330\\
\noindent Dolphin, A.~E., Walker, 
A.~R., Hodge, P.~W., Mateo, M., Olszewski, E.~W., Schommer, R.~A., \& 
Suntzeff, N.~B.\ 2001, \apj, 562, 303 \\
\noindent Dopita, M.~A., Lawrence, 
C.~J., Ford, H.~C., \& Webster, B.~L.\ 1985, \apj, 296, 390\\
\noindent Fellhauer, M. \etal 2006 astro-ph/0605026\\
\noindent Font, A.~S., Johnston, 
K.~V., Bullock, J.~S., \& Robertson, B.~E.\ 2006, \apj, 638, 585\\
\noindent Freedman, W.~L., et al.\ 2001, \apj, 553, 47\\
\noindent Freire, P.~C., Camilo, 
F., Kramer, M., Lorimer, D.~R., Lyne, A.~G., Manchester, R.~N., \& D'Amico, 
N.\ 2003, \mnras, 340, 1359 \\
\noindent Gardiner, L.~T., \& Noguchi, M.\ 1996, \mnras, 278, 191 (GN96)\\
\noindent Gardiner, L.~T., Sawa, T., \& Fujimoto, M.\ 1994, \mnras, 266, 567\\
\noindent Geha, M.~et al. 2003, AJ, 125, 1\\
\noindent Haberl, F., Filipovi{\'c}, M.~D., Pietsch, W., \& Kahabka, 
P.\ 2000, \aaps, 142, 41 \\
\noindent Hardy, E., Suntzeff, N.~B., \& Azzopardi, M.\ 1989, \apj, 344, 210\\
\noindent Harris, J., \& Zaritsky, D.\ 2006, \aj, 131, 2514\\ 
\noindent Harris, J., \& Zaritsky, D.\ 2001, \apjs, 136, 25\\ 
\noindent Hatzidimitriou, 
D., Cannon, R.~D., \& Hawkins, M.~R.~S.\ 1993, \mnras, 261, 873 \\
\noindent Heller, P.~\& Rohlfs, K. 1994, A\&A, 291, 743\\
\noindent Helmi, A.\ 2004, \apjl, 610, L97 \\
\noindent Hernquist, L.\ 1990, \apj, 356, 359 \\
\noindent Holtzman, J.~A., et al.\ 1997, \aj, 113, 656\\ 
\noindent Ibata, R.~A., Gilmore, G., \& Irwin, M.~J.\ 1994, \nat, 370, 194\\
\noindent Irwin, M.\ 1999, IAU Symposium, 192, 409\\ 
\noindent Irwin, M., Demers, S., \& 
Kunkel, W.\ 1996, Bulletin of the American Astronomical Society, 28, 932\\
\noindent  Johnston, K.~V., Law, 
D.~R., \& Majewski, S.~R.\ 2005, \apj, 619, 800\\ 
\noindent Johnston, K.~V., Zhao, 
H., Spergel, D.~N., \& Hernquist, L.\ 1999, \apjl, 512, L109\\
\noindent Kallivayalil, N., 
van der Marel, R.~P., Alcock, C., Axelrod, T., Cook, K.~H., Drake, A.~J., 
\& Geha, M.\ 2006, \apj, 638, 772 \\
\noindent Krist, J.\ 2003, ACS Instrument Science Report 03-06 
(Baltimore: Space Telescope Science Institute)\\
\noindent Kroupa, P.~\& Bastian, U. 1997, New Astronomy, 2, 77\\
\noindent Law, D.~R., Johnston, 
K.~V., \& Majewski, S.~R.\ 2005, \apj, 619, 807 \\
\noindent Lin, D.~N.~C., \& Lynden-Bell, D.\ 1982, \mnras, 198, 707\\ 
\noindent Lin, D.~N.~C., Jones, B.~F., \& Klemola, A.~R. 1995, ApJ, 439, 652\\
\noindent Majewski, S.~R., 
Skrutskie, M.~F., Weinberg, M.~D., \& Ostheimer, J.~C.\ 2003, \apj, 599, 
1082 \\
\noindent Mastropietro, C., 
Moore, B., Mayer, L., Wadsley, J., \& Stadel, J.\ 2005, \mnras, 363, 509\\
\noindent Mathewson, D.~S., 
Ford, V.~L., \& Visvanathan, N.\ 1986, \apj, 301, 664 \\
\noindent Mathewson, D.~S., 
Cleary, M.~N., \& Murray, J.~D.\ 1974, \apj, 190, 291 \\
\noindent Momany, Y., \& Zaggia, S.\ 2005, \aap, 437, 339\\ 
\noindent Moore, B., \& Davis, M.\ 1994, \mnras, 270, 209 \\
\noindent Mu{\~n}oz \etal 2006 astro-ph/0605098\\
\noindent Murai, T., \& Fujimoto, M.\ 1980, \pasj, 32, 581\\
\noindent Navarro, J.~F., Frenk, 
C.~S., \& White, S.~D.~M.\ 1997, \apj, 490, 493 \\
\noindent Pe{\~n}arrubia, J., et al.\ 2005, astro-ph/0512507\\
\noindent Piatti, A.~E., 
Sarajedini, A., Geisler, D., Bica, E., \& Clari{\'a}, J.~J.\ 2002, \mnras, 
329, 556 \\
\noindent Putman, M.~E., et al.\ 1998, \nat, 394, 752 \\
\noindent Putman, M.~E., 
Staveley-Smith, L., Freeman, K.~C., Gibson, B.~K., \& Barnes, D.~G.\ 2003, 
\apj, 586, 170 \\
\noindent Schommer, R.~A., 
Suntzeff, N.~B., Olszewski, E.~W., \& Harris, H.~C.\ 1992, \aj, 103, 447\\
\noindent Smecker-Hane, 
T.~A., Cole, A.~A., Gallagher, J.~S., III, \& Stetson, P.~B.\ 2002, \apj, 
566, 239 \\
\noindent Springel, V., Yoshida, 
N., \& White, S.~D.~M.\ 2001, New Astronomy, 6, 79 \\
\noindent Stanimirovi{\'c}, S., Staveley-Smith, L., \& Jones, 
P.~A.\ 2004, \apj, 604, 176 \\
\noindent  van den Bergh, S.\ 2000, 
The galaxies of the Local Group, Cambridge, UK: Cambridge University Press\\
\noindent van der Marel, R.~P., Alves, D.~R., Hardy, E., \& Suntzeff, N.~B.\
2002, AJ, 124, 2639\\
\noindent Wannier, P., \& Wrixon, G.~T.\ 1972, \apjl, 173, L119\\ 
\noindent Wayte, S.~R.\ 1991, IAU Symp.~148: The Magellanic Clouds, 148, 447 \\
\noindent Welch, D.~L., McLaren, 
R.~A., Madore, B.~F., \& McAlary, C.~W.\ 1987, \apj, 321, 162 \\
\noindent Weinberg, M.~D., \& Blitz, L.\ 2006, \apjl, 641, L33\\ 
\noindent Weinberg, M.~D.\ 2000, \apj, 532, 922\\ 
\noindent Westerlund, B.~E. \ 1997, The Magellanic Clouds (Cambridge:
Cambridge Univ. Press)\\
\noindent Whitmore, B.~C.\ 1999, IAU 
Symp.~186: Galaxy Interactions at Low and High Redshift, 186, 251 \\
\noindent Yoshizawa, A.~M., \& Noguchi, M.\ 2003, \mnras, 339, 1135 \\
\noindent Zentner, A.~R., \& Bullock, J.~S.\ 2003, \apj, 598, 49\\
 
\begin{deluxetable}{llllcccccccccccc}
\tabletypesize{\tiny}
\rotate
\tablewidth{0pt}
\tablecolumns{16}
\tablecaption{Sample and Observations}
\tablehead{
\colhead{ID}  & \colhead{QSOname} & 
\colhead{RA} & \colhead{DEC} &
\colhead{$V$} & \colhead{$z$} & \multicolumn{5}{c}{epoch1} 
& \multicolumn{4}{c}{epoch2} & \colhead{$\Delta$time}}
\startdata
 &  &  &  &  &  & date & visit 
& $T_{exp}$ & $T_{exp}$ & 
ORIENTAT & 
date & visit & $T_{exp}$ & ORIENTAT &   \\
  &   &  &  &  &  &  &  
& F606W & F814W &  &  &  &   F606W &  &   \\
 & & (H,M,S) & (deg, ', '')     &  &  &  &  & 
(min) & (min) & (deg) &  &  & (min) & (deg) & (yrs)\\
S1 & 208.16034.100 & 00 51 17.0 &  -72 16 51.3 & 18.6 & 0.49 & 2002-08-27
& 14 & 6.7 & 1.7 & 140.8 & 2004-07-16 & 13 & 9.5 & 95.9 & 1.9\\
S2 & 207.16316.446 &  00 55 34.7  &  -72 28 33.9	& 18.9 & 0.56 &
2002-08-26 & 15 & 6.7 & 1.7 &  137.4 & 2004-07-17 & 14 & 9.6 & 96.1 & 1.9 \\
S3 & 211.16703.311 &   01 02 14.5 &  -73 16 26.6 & 20.3 & 2.18 &
2002-09-03 & 16 & 6.7 & 1.7 & 129.6 & 2004-07-13 & 15 & 16 & 90.8 & 1.9\\
S4 & QJ0036-7227  &  00 36 39.7 &  -72 27 42.0 & 19.6  & 1.62 &
2002-09-13 & 17 & 8 & 1.7 & 160.9 & 2005-06-19 & 16 & 18 &  72.1 & 2.8\\
S5 & 211.16765.212 &  01 02 34.7 & -72 54 23.8 &	18.4 & 2.13 &
2003-02-15 & 36 & 6.7 & 1.7 &  -51.1 & 2004-07-15 & 33 & 6.7 & 91.5 & 1.4\\
\enddata
\tablecomments{Column 1 is a field identification. Column 2 gives the
  MACHO ID for referencing with Geha \etal (2003). Columns 3 \& 4 give
  the RA and Dec of the QSOs (J2000). The $V$ magnitudes
quoted in column 5 are from our $HST$ data. Column 6 gives the QSO
redshifts. S4 is a ROSAT object from Haberl \etal (2000). 
All other redshifts are from Geha \etal (2003). Epoch 1 
has program ID 9046 (Cycle 11) and epoch 2 has program ID 10130 (Cycle
13). The date of observation, $HST$ phase II visit number, and
exposure time for each field are listed.  
ORIENTAT is the position angle on the sky of the detector $y$-axis (in 
degrees east of north), and $\Delta$time gives the baseline.}
\end{deluxetable}

\pagebreak

\begin{deluxetable}{lcccccccc}
\tabletypesize{\scriptsize}
\tablewidth{0pt}
\tablecolumns{9}
\tablecaption{Results}
\tablehead{
\colhead{ID}  & \colhead{$N_{\rm{sources}}$} & 
\colhead{$N_{\rm{master}}$} & \colhead{$N_{\rm{used}}$\tablenotemark{a}} &
\colhead{$\mu_N$} & \colhead{$\mu_W$} & \colhead{$\delta\mu_N$} 
& \colhead{$\delta\mu_W$} & \colhead{Used?}}
\startdata
   &   &    &    & ($\masyr$)  & ($\masyr$)  & ($\masyr$)  & ($\masyr$)  &  \\
S1 & 247 & 71 & 32 & -1.136 & -0.860 & 0.095  & 0.113 & Yes\\
S2 & 303 & 117 & 54 & -1.208  & -0.825  &   0.076 &  0.073 & Yes\\
S3 & 235 & 87 & 45 & -1.201 &  -1.022 &  0.109  & 0.091 & Yes\\
S4 & 68 & 10 & 4 & -0.866 & -0.303 &  0.177 & 0.073  & No\\
S5 & 242 & 100 & 42 &  -1.143 &  -1.471 & 0.130  & 0.108 & Yes \\
\enddata
\tablenotetext{a}{$N_{\rm{sources}}$ refers to the number of real sources 
(detected in at least half of the images). $N_{\rm{master}}$ refers to the 
number or sources in the master-list, i.e. detected in every image in 
every epoch. $N_{\rm{used}}$ refers to the number of sources 
that are used in the final linear transformations after the PM and 
$\delta$PM cuts. Columns 5-8 contain the PM estimates and their 
errors for each field. The last column notes if the particular 
field was used in our final estimate of the center of mass motion 
of the SMC (Equation~(\ref{PMfinal})).} 
\end{deluxetable}

\pagebreak

\begin{deluxetable}{llll}
\tabletypesize{\scriptsize}
\rotate
\tablewidth{0pt}
\tablecolumns{4}
\tablecaption{Orbital Parameters}
\tablehead{
\colhead{}  & \colhead{LMC} & 
\colhead{SMC} & \colhead{References}}
\startdata
Line-of-sight velocity ($\kms$) & $262.1\pm3.4$ & $146\pm0.6$ & van
der Marel \etal 2002; Harris \& Zaritsky 2006\\
Proper Motions ($W$,$N$) ($\masyr$) & $-2.03\pm0.08$, $0.44\pm0.05$ &
$-1.16\pm0.18$, $-1.17\pm0.18$ & Paper~I; this work\\
Distance Moduli & $18.50\pm0.1$ & $18.95\pm0.1$ & Freedman \etal 2001;
Cioni \etal 2000\\
Current positions ($\alpha$, $\delta$) (deg) & $81.9\pm 0.3$, 
$-69.9\pm0.3$ & $13.2\pm0.3$,$-72.5\pm0.3$ & van der Marel \etal 2002;
Westerlund 1997; Stanimirovi{\'c} \etal 2004\\
Galactic Coordinates ($l$, $b$) & $280.5$, $-32.5$ & $302.8$, $-44.6$
& -- \\
Current positions ($X,Y,Z$) (kpc) & $-0.8, -41.5, -26.9$ & $15.3, 
-36.9, -43.3$ & -- \\
Space velocities ($v_X, v_Y, v_Z$) ($\kms$) & $-86 \pm 12, 
-268 \pm 11, 252 \pm 16$, & $-87 \pm 48, -247 \pm 42, 149 \pm 37$ & -- \\
Galactocentric radial velocities ($\kms$) & $89 \pm 4$ & 
$23 \pm 7$ & -- \\
Galactocentric tangential velocities ($\kms$) & $367 \pm 18$ & 
$301 \pm 52$ & -- \\
\enddata
\tablecomments{Positions and velocities of the Clouds as discussed in
  the text and Paper~I. The last column lists the sources of the
  adopted values. No source is listed for the values in the bottom 5
  lines because they follow uniquely from the top 4 lines.}
\end{deluxetable}

\end{document}